\documentclass[usenatbib]{mnras}
\pdfoutput=1
\usepackage{mathptmx}
\usepackage{txfonts}

\usepackage[T1]{fontenc}
\usepackage{ae,aecompl}

\usepackage{graphicx}
\usepackage[utf8]{inputenc}
\usepackage[english]{babel}

\newcommand{\Ms}{\ensuremath{M_{\odot}}}
\newcommand{\Zs}{\ensuremath{Z_{\odot}}}
\newcommand{\eg}{{\it e.g.}}
\newcommand{\cf}{{\it c.f.~}}
\newcommand{\ie}{{\it i.e.}}

\newcommand{\beq}{\begin{equation}}
\newcommand{\eeq}{\end{equation}}
\newcommand{\mtot}{\ensuremath{M_{\rm tot}}}

\newcommand{\kmps}{\ensuremath{{\rm~km~s}^{-1}}}
\newcommand{\peryr}{\ensuremath{{\rm~yr}^{-1}}}

\newcommand{\rhocl}{\ensuremath{\rho_{cl}}}
\newcommand{\mcl}{\ensuremath{M_{cl}}}
\newcommand{\rh}{\ensuremath{r_h}}
\newcommand{\rligo}{\ensuremath{{\rm R}_{\rm LIGO}}}
\newcommand{\rhbh}{\ensuremath{r_{\rm h,BH}}}
\newcommand{\rhns}{\ensuremath{r_{\rm h,NS}}}

\newcommand{\tmrg}{\ensuremath{t_{\rm mrg}}}
\newcommand{\taumrg}{\ensuremath{\tau_{\rm mrg}}}
\newcommand{\tej}{\ensuremath{t_{\rm ej}}}

\newcommand{\nmrgin}{\ensuremath{N_{\rm mrg,in}}}
\newcommand{\nmrgout}{\ensuremath{N_{\rm mrg,out}}}
\newcommand{\nbhbound}{\ensuremath{N_{\rm BH,bound}}}
\newcommand{\nbhboundmx}{\ensuremath{N_{\rm BH,bound,max}}}
\newcommand{\nbseven}{{\tt NBODY7~}}
\newcommand{\nbsix}{{\tt NBODY6 }}
\newcommand{\bse}{{\tt BSE }}
\newcommand{\sse}{{\tt SSE }}
\title[Stellar-mass black holes in open clusters]
{Stellar-mass black holes in young massive and open stellar
clusters and their role in gravitational-wave generation}

\author[S. Banerjee]{
Sambaran Banerjee$^{1,2}$\thanks{E-mail: sambaran@astro.uni-bonn.de (SB)}
\\
$^{1}$Argelander-Institut f\"ur Astronomie (AIfA),
Auf dem H\"ugel 71, D-53121, Bonn, Germany\\
$^{2}$Helmholtz-Instituts f\"ur Strahlen- und Kernphysik (HISKP),
Nussallee 14-16, D-53115 Bonn, Germany
}

\pubyear{2016}

\begin{document}
\label{firstpage}
\pagerange{\pageref{firstpage}--\pageref{lastpage}} 
\maketitle

\begin{abstract}
Stellar-remnant black holes (BH) in dense stellar clusters have always drawn attention due to their potential
in a number of phenomena, especially
the dynamical formation of binary black holes (BBH), which potentially coalesce
via gravitational-wave (GW) radiation.
This study presents a preliminary set of evolutionary models of compact stellar clusters
with initial masses ranging over $1.0\times10^4\Ms-5.0\times10^4\Ms$, and
half-mass radius of 2 or 1 pc, that is typical for young massive and starburst clusters.
They have metallicities between $0.05\Zs-\Zs$.
Including contemporary schemes for stellar wind and remnant formation,
such model clusters are evolved, for the first time, using the state-of-the-art
direct N-body evolution program \nbseven, until their dissolution or at least for 10 Gyr.
That way, a self-regulatory behaviour in the effects of dynamical interactions among the BHs
is demonstrated. In contrast to earlier studies, the BBH coalescences obtained
in these models show a prominence in triple-mediated coalescences while
being bound to the clusters, compared to those occurring among the BBHs that are dynamically
ejected from the clusters. A broader mass spectrum of the BHs and lower escape velocities
of the clusters explored here might cause this difference,
which is yet to be fully understood. Among the BBH coalescences obtained here,
there are ones that resemble the detected GW151226, LVT151012, and
GW150914 events and also ones which are even more massive. A preliminary
estimate suggests few 10s-100s of BBH coalescences per year, originating
due to dynamics in stellar clusters, that can be detected by the LIGO at its
design sensitivity.
\end{abstract}

\begin{keywords}
open clusters and associations: general -- globular clusters: general --
stars: kinematics and dynamics -- stars: black holes -- methods: numerical -- 
gravitational waves
\end{keywords}

\section{Introduction}\label{intro}

The study of dynamical interactions of black holes (hereafter BH)
in dense stellar systems is now
nearly 30 years old. A key point of interest in this topic has always been
the possibility of the generation of gravitational waves (hereafter GW) from
dynamically-formed binary black holes (hereafter BBH). The interest in the topic
has naturally got rejuvenated right after the first-time detection of GW from
two BBH inspiral events by the Advanced LIGO detector, namely,
the GW150914 \citep{2016PhRvL.116f1102A,2016ApJ...818L..22A}, the GW151226 \citep{2016PhRvL.116x1103A}
and the marginal detection event LVT151012 \citep{2016PhRvL.116f1102A}.
Generally speaking, for BBHs composed of stellar-remnant BHs such as the
above detected ones, which are typically of $\sim10\Ms-\sim100\Ms$, the frequency
of the emitted GW during their inspiral phase falls within the LIGO's detection band \citep{2016PhRvL.116f1102A},
placing such BBHs among the most promising sources for the LIGO.
The bottom-line scenario of formation of dynamical (stellar-mass) BBHs
in star clusters is straightforward: if a certain number of BHs receive sufficiently low
natal kicks, during their formation via core-collapse (supernovae) of massive
stars, that they remain bound to the gravitational potential of the cluster,
they would segregate to the innermost regions of the cluster. Depending on
the number of BHs retained in the cluster and their masses
(which is $\sim10$ to $\sim100$ times the average stellar mass depending
on their progenitor stars' wind; see below), the system
of bound BHs might undergo a runaway mass segregation (mass-stratification
or Spitzer instability; \citealt{1987degc.book.....S}) to form
a central and highly dense subsystem of BHs, where they would continuously
interact. Otherwise, the dynamical friction of the dense stellar
background would as well tend to keep the BHs centrally concentrated
(as these BHs are much more massive than the normal stars, they
would segregate towards the cluster's center simply due to dynamical
friction, rather than being driven by two-body relaxation).

Such a dense BH-core serves as a constant resource for dynamically
forming BBH, mainly via the three-body mechanism \citep{1987degc.book.....S,2003gmbp.book.....H}.
The subsequent frequent and super-elastic \citep{1987degc.book.....S}
dynamical encounters of a BBH with other single
BHs and BBHs serve as a recipe for (a) injecting kinetic energy (K.E.)
into the BH sub-cluster and as well into the whole star cluster causing the
latter to expand \citep{2007MNRAS.379L..40M,2008MNRAS.386...65M},
(b) ejecting single and binary BHs from the cluster depleting the
BH population and (c) forming triple-BH systems within the BH-core. The ejected
BBHs are typically dynamically tightened (hardened; \citealt{1975MNRAS.173..729H})
and also eccentric, an adequate combination of which \citep{Peters_1964} would lead to the inspiral
via GW radiation and the coalescence of a BBH within the Hubble time.
The triple-BHs that are bound to the clusters, on the other hand,
would typically undergo Kozai-Lidov oscillations \citep{Kozai_1962}
leading to large eccentricity boost and hence GW inspiral and coalescence
of the inner binary, provided this happens
before the triple gets perturbed by an intruder. In other words, although a star cluster
continues to eject single and binary BHs and form BH-triples until its BH reservoir is (nearly) depleted,
the occurrence of a dynamical BBH inspiral is a probabilistic phenomenon.

Beginning from 1990s, aspects of the above mechanism is studied at various
levels of detail. Following preliminary but pioneering studies such as
\citet{1993Natur.364..421K,1993Natur.364..423S,2000ApJ...528L..17P},
more recent direct N-body (\eg, \citealt{2010MNRAS.402..371B,2012MNRAS.422..841A,2013MNRAS.430L..30S,Wang_2016,2016PASA...33...36H})
and Monte-Carlo (\eg, \citealt{Downing_2010,Downing_2011,Rodriguez_2015,Morscher_2015,Rodriguez_2016,Rodriguez_2016a,2016arXiv160300884C,2016arXiv160906689C,Askar_2016})
calculations of model stellar clusters study the dynamically-driven depletion of
BHs, the resulting feedback onto the cluster and the BBH inspirals
self consistently and in much more detail. Adopting somewhat simpler
but realistic conditions, detailed semi-analytic studies of these aspects
have also been performed recently \citep{Breen_2013,Breen_2013a,ArcaSedda_2016}.
Also, such a semi-analytic modelling, in the context of $\sim10^7\Ms$ nuclear stellar clusters
and with focus on stellar-mass BBH inspirals produced by them,
has been recently performed by \citet{Antonini_2016}.
By nature, Monte-Carlo calculations
are restricted to massive clusters, initially $\sim 10^5\Ms - 10^6\Ms$, that
are more representatives (or progenitors) of classical globular
clusters (hereafter GC), although young clusters of up to $\sim 10^7\Ms$ are
observed in nearby starburst galaxies (\eg, in the Antennae
\citealt{Larsen_2009,Johnson_2015}) and
the impact on BBH inspiral rate due to age-spread among massive clusters has been
explored very recently \citep{2016arXiv160906689C}. The overall conclusion of such studies
is that the dynamical BBH inspiral rate is $\approx 5-10 {\rm~yr}^{-1} {\rm~Gpc}^{-3}$,
which would contribute to the detection of several 10s of BBH inspriral per year
with the Advanced LIGO, given its proposed full sensitivity \citep{2010MNRAS.402..371B,Rodriguez_2016}.
These studies also infer that a GW150914-like BBH coalescence is intrinsically
rare to be produced from a star cluster compared to the other two events
\citep{Rodriguez_2016a,2016arXiv160906689C}.

Another important corollary is
that the BH population in massive clusters, although decays monotonically with
time due to dynamical interactions, it hardly gets completely depleted
even in a Hubble time, so that a substantial population of BHs would retain
even in old globular clusters. In a nutshell, this is due to the
energy generation via dynamical encounters in the BH-core that results
in significant expansion of both the parent cluster and the BH-core itself,
suppressing the BH-BH interaction rate. This ``self-regulation'' causes
the BH population to decline but exponentially; see below and also
\citet{Morscher_2015}. This is consistent with the recent identification
of stellar-mass BH candidates in the Galactic globular clusters M22 and 47 Tuc
\citep{Strader_2012,Miller_Jones_2015}.

The present work focuses on the other end of the problem, namely,
the role of intermediate-mass and open stellar clusters in generating
BBH inspirals. In this paper, by intermediate-mass clusters, we will imply
those within the mass range $10^4\Ms - 10^5\Ms$, the young ($\lesssim100$ Myr) versions of which
are popularly called ``young massive clusters'' (YMCs) and ``starburst clusters''
(when $\lesssim4$ Myr old; such young clusters with $>10^5\Ms$ are often denoted as
``super star clusters''). Clusters of $<10^4\Ms$ will be called
open clusters; as such, there is no strict boundaries defined in
the literature between these different ``types'' of clusters. A reliable
and self-consistent evolutionary modelling of clusters of such masses with
realistic ingredients is possible only by direct N-body integration. Monte-Carlo calculations already
become unsatisfactory for the corresponding typical particle numbers
($N\lesssim1.7\times10^5$), due to larger statistical fluctuations and
various timescales becoming closer to the cluster's
overall two-body relaxation time.

After a recess of about a decade since a few notable initial
studies on this topic \citep{1993Natur.364..421K,1993Natur.364..423S,2000ApJ...528L..17P},
\citet{2010MNRAS.402..371B} have, for the first time, investigated the
dynamical behaviour of a population of $\approx10\Ms$ BHs in intermediate-mass
compact (initial half-mass radius 1-2 pc) stellar clusters, where the
dynamics of the BHs have been treated self-consistently using direct
N-body calculations. As such, the above work is the first of its kind where
the dynamics of the BHs in stellar clusters and the consequent
impact on the cluster and the production of BBH coalescences have
been studied explicitly. However, soon after,
both observations of BH candidates
in sub-solar metallicity regions in the Local Group and theoretical studies
of the mass distribution of stellar remnants based on revised
wind prescriptions for high-mass stars have suggested that
stellar-remnant BHs can, in fact, be much more massive than
the contemporarily accepted $\approx10\Ms$ BHs, especially at low abundances
\citep{Belczynski_2010}. Of course, the existence of such stellar
BHs is now confirmed after the GW150914 event which involves
$\approx30\Ms$ BHs (also by the LVT151012 event).
Given the current knowledge on stellar winds in
different evolutionary stages of massive stars as a function of metallicity,
determining the remnant BH mass, it is likely that such $\approx30\Ms$ BHs
are formed in sub-solar metallicity regions \citep{Belczynski_2010,Spera_2015}.

Given that nearly all of the N-body studies of long-term BH dynamics in
stellar clusters so far assumes $\approx10\Ms$ BHs that is
representative of solar-like metallicity \citep{2010MNRAS.402..371B,2012MNRAS.422..841A,2013MNRAS.430L..30S,Wang_2016},
it is now undoubtedly worthwhile to revisit the problem with revised BH masses.
Especially, at lower metallicities, not only the stellar-remnant BHs would be
substantially more massive, but also they would have a wider mass spectrum.
This, in turn, would influence the nature of the dynamical interactions
in a star cluster's ``BH-engine'' and hence the latter's impact on the cluster and
on the BBH coalescences from it, as we will see in the following sections.
Since direct N-body integration (without force softening) tracks
all sorts of dynamical encounters, and particularly the close ones, with
full consistency and without any assumptions, it is the ideal approach
for this study. In the present work, it is for the first time that
model parsec-scale intermediate-mass and open star clusters
of varying metallicities are evolved
using direct N-body integration from their zero-age until the Hubble time
(at least for 10 Gyr). The {\tt NBODY7} code, with
modified stellar mass loss and remnant formation prescriptions
adopting those of \citet{Belczynski_2010} are used for this purpose;
that way the BH-dynamics is studied as consistently and realistically
as possible now. Given the long computing times, the present set of models cover
the relevant parameter space only preliminarily and thus provide
limited statistics. However, they would still comprise the state-of-the-art
set of N-body calculations already suggesting intriguing and new conclusions,
as we shall see in the following sections. Recently, lower-mass clusters
of initially $\sim10^3\Ms$ are evolved, using the direct N-body method containing similar
model ingredients as in here (Sec.~\ref{newwind}), for shorter evolutionary times of $\sim100$ Myr, to
study the dynamical interactions involving BHs over
young ages and low metallicities \citep{Mapelli_2013,Ziosi_2014}. 
It is also worth recalling the ``Dragon Simulations''
\citep{Wang_2016} in this context, where relatively extended, $\approx3-8$ pc-sized,
much more massive clusters of $N\approx10^6$ stars (they can also be taken as representatives
of galactic nuclear clusters) are evolved for nearly a Hubble time
and the properties of the BH population are studied,
through direct N-body calculations using the {\tt NBODY6++} program
\citep{Wang_2015}.

This paper is organized as follows: in Sec.~\ref{calc}, the model calculations
are described (in Sec.~\ref{runs}) following a brief introduction to
the \nbseven code (Sec.~\ref{nbprog}) and
its modifications for the present purpose (Sec.~\ref{newwind}). In Sec.~\ref{result},
the results of these calculations are discussed, with focus on the
general dynamical behaviour of BHs and its impact on the parent cluster (Sec.~\ref{bhdyn}),
and as well on the dynamical production of BBHs and their
coalescences via GW radiation (Sec.~\ref{bhbin}). These coalescences are
compared with those detected by the LIGO (Sec.~\ref{LIGOcomp}) and a preliminary
estimate of the LIGO detection rate of dynamically-generated BBH mergers
is suggested (Sec.~\ref{LIGOrate}). The inferences are summarized
in Sec.~\ref{conclude}.

\begin{figure}
\includegraphics[width=8.0cm,angle=0]{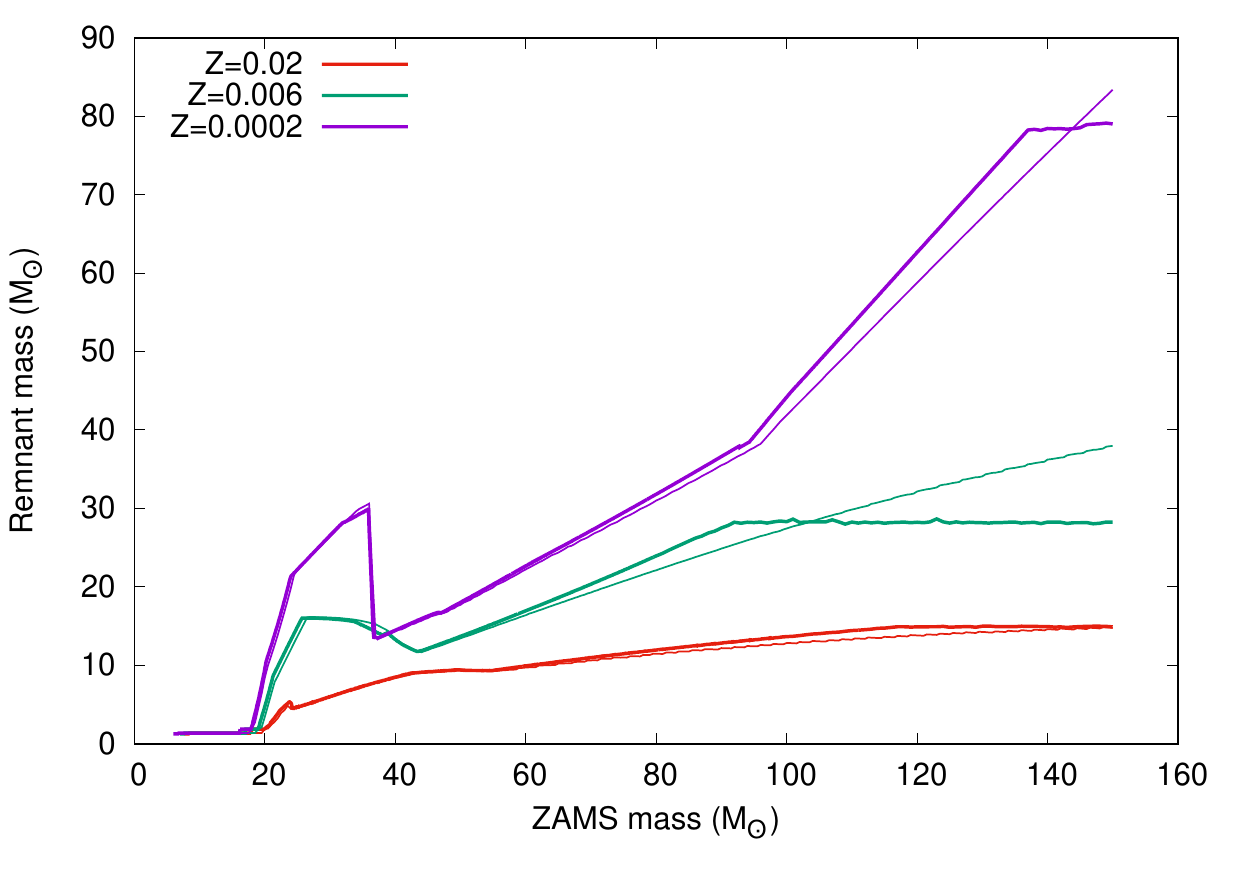}
\includegraphics[width=8.0cm,angle=0]{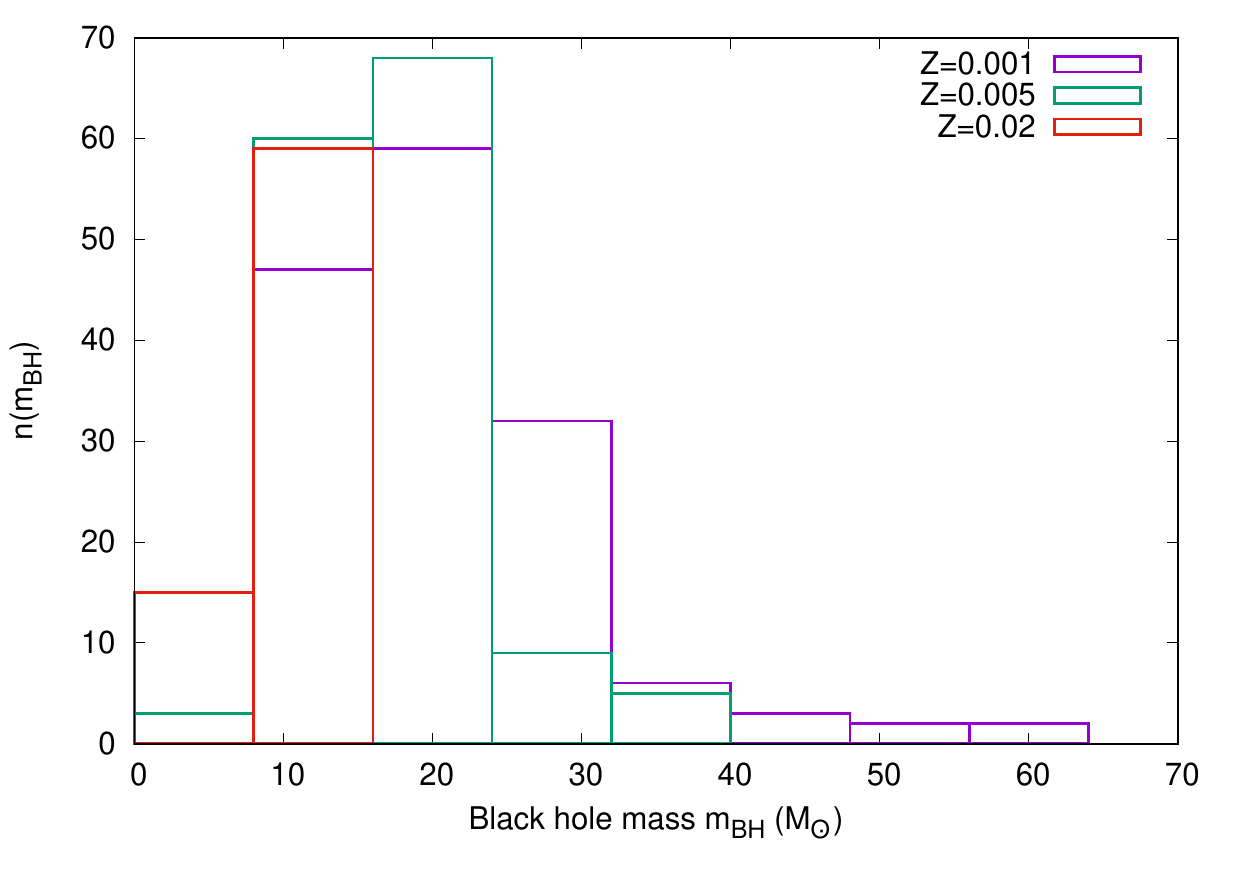}
\caption{{\bf Top:} Remnant mass as a function of zero-age main sequence (ZAMS) mass
	for the \citet{Belczynski_2008,Belczynski_2010} stellar wind and remnant formation
	schemes adopted in this study, which is obtained from the variant of
	\bse that is integrated with \nbseven (thin lines). They agree reasonably
	well with those from B10 (thick lines) for the respective metallicities
	($Z=\Zs,\Zs/4,\Zs/100$; see B10),
	which have been obtained using their {\tt StarTrack} program.
	{\bf Bottom:} Mass distributions of BHs that remain bound to the
	$\mcl(0)=5\times10^4\Ms$ computed model clusters (Table~\ref{tab:comp}), right after their
	formation, for $Z=\Zs,\Zs/4,\Zs/20$.}
\label{fig:bhmass}
\end{figure}

\begin{figure*}
\centering
\includegraphics[width=8.0cm,angle=0]{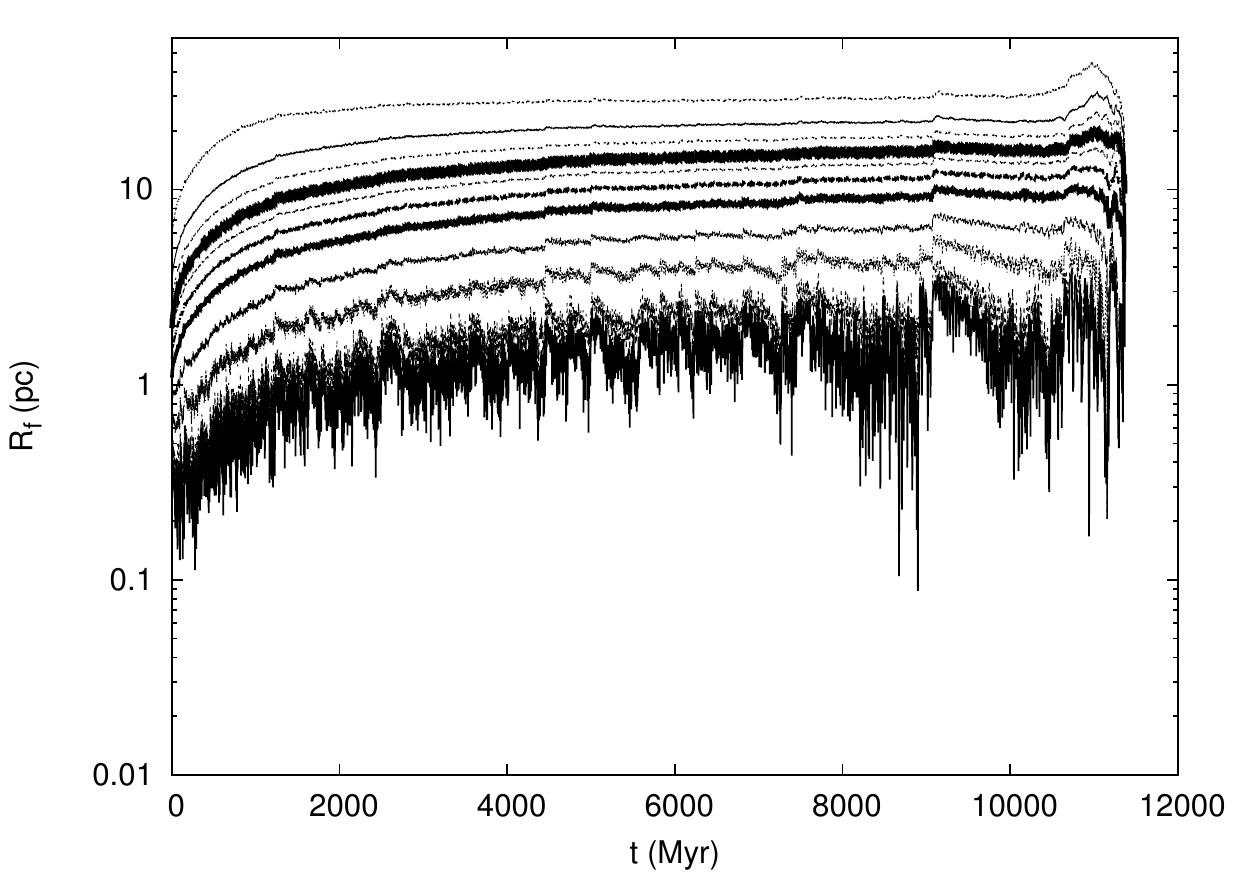}
\includegraphics[width=8.0cm,angle=0]{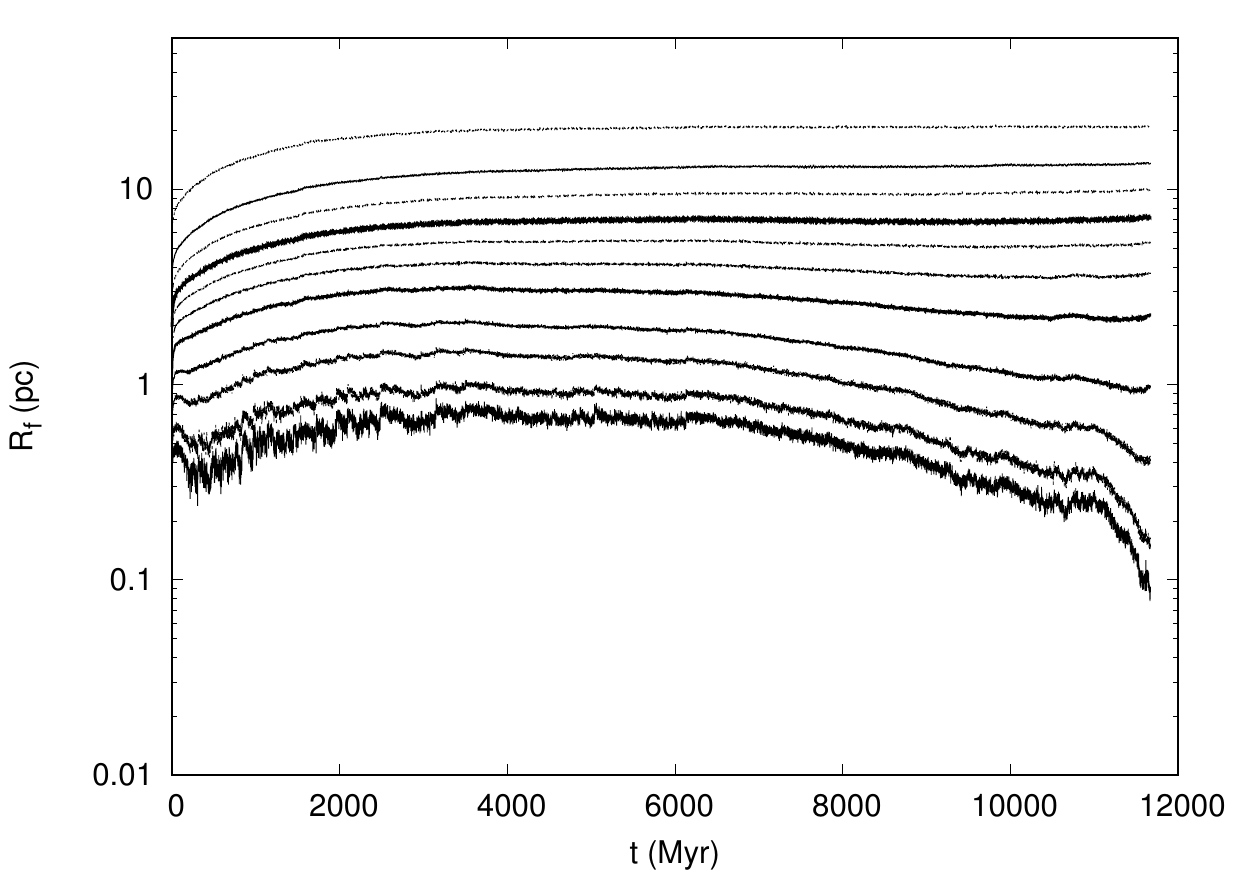}
\includegraphics[width=8.0cm,angle=0]{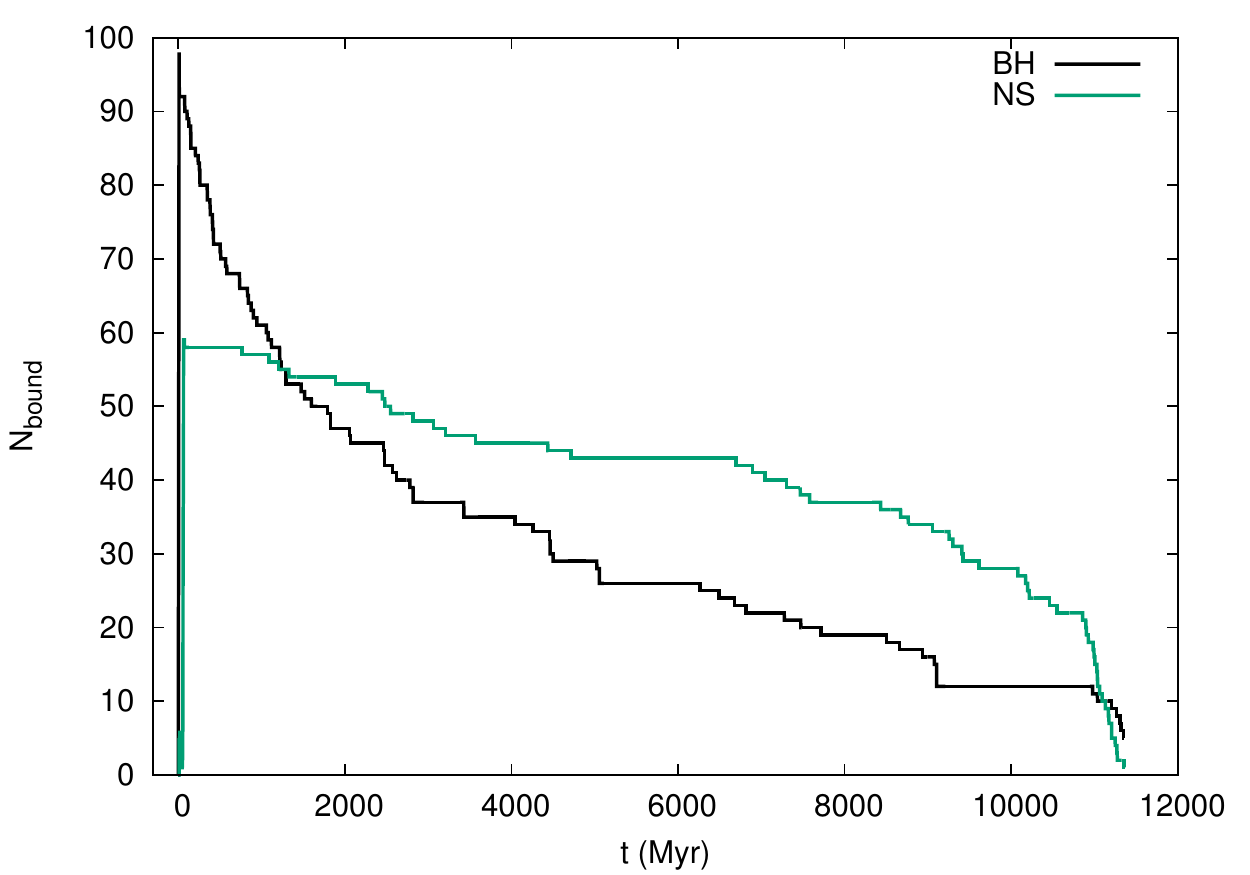}
\includegraphics[width=8.0cm,angle=0]{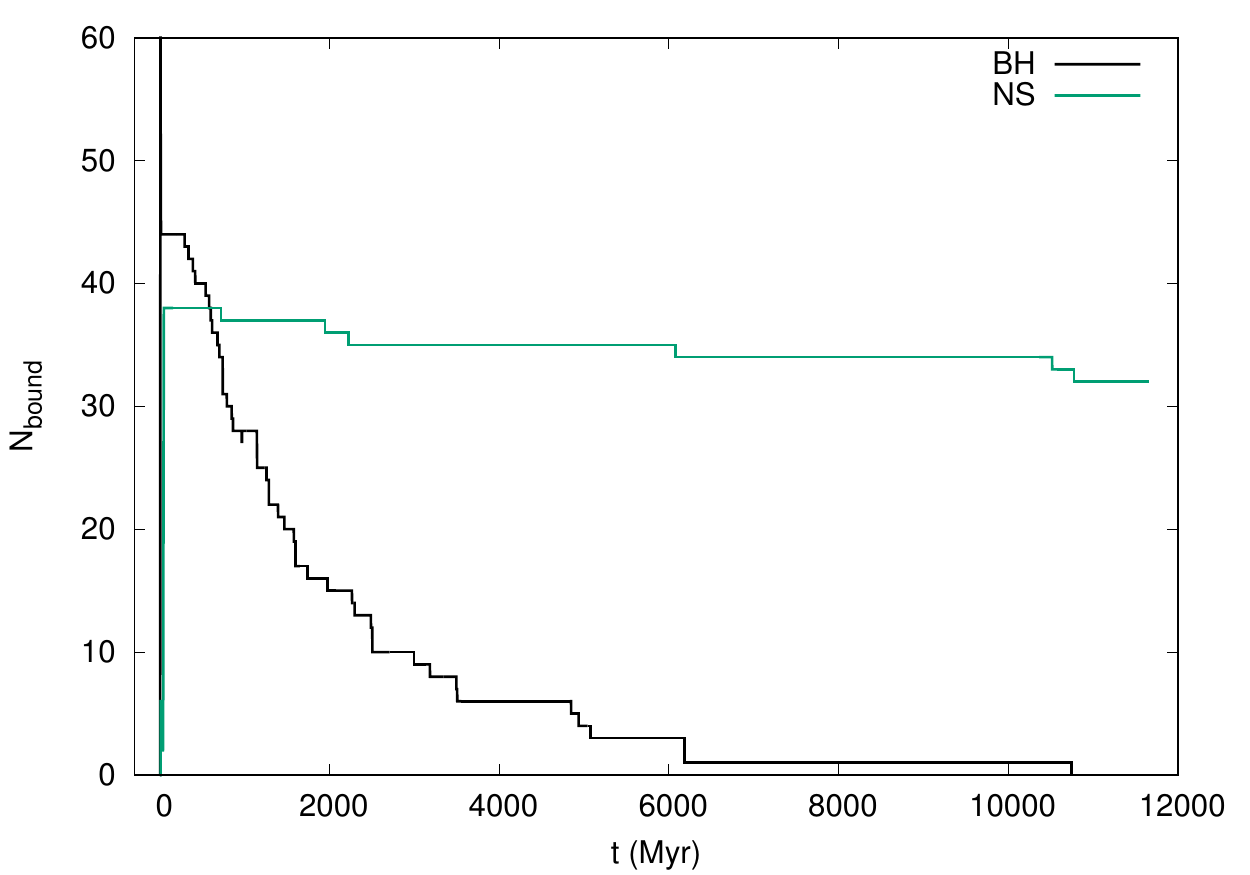}
\includegraphics[width=8.0cm,angle=0]{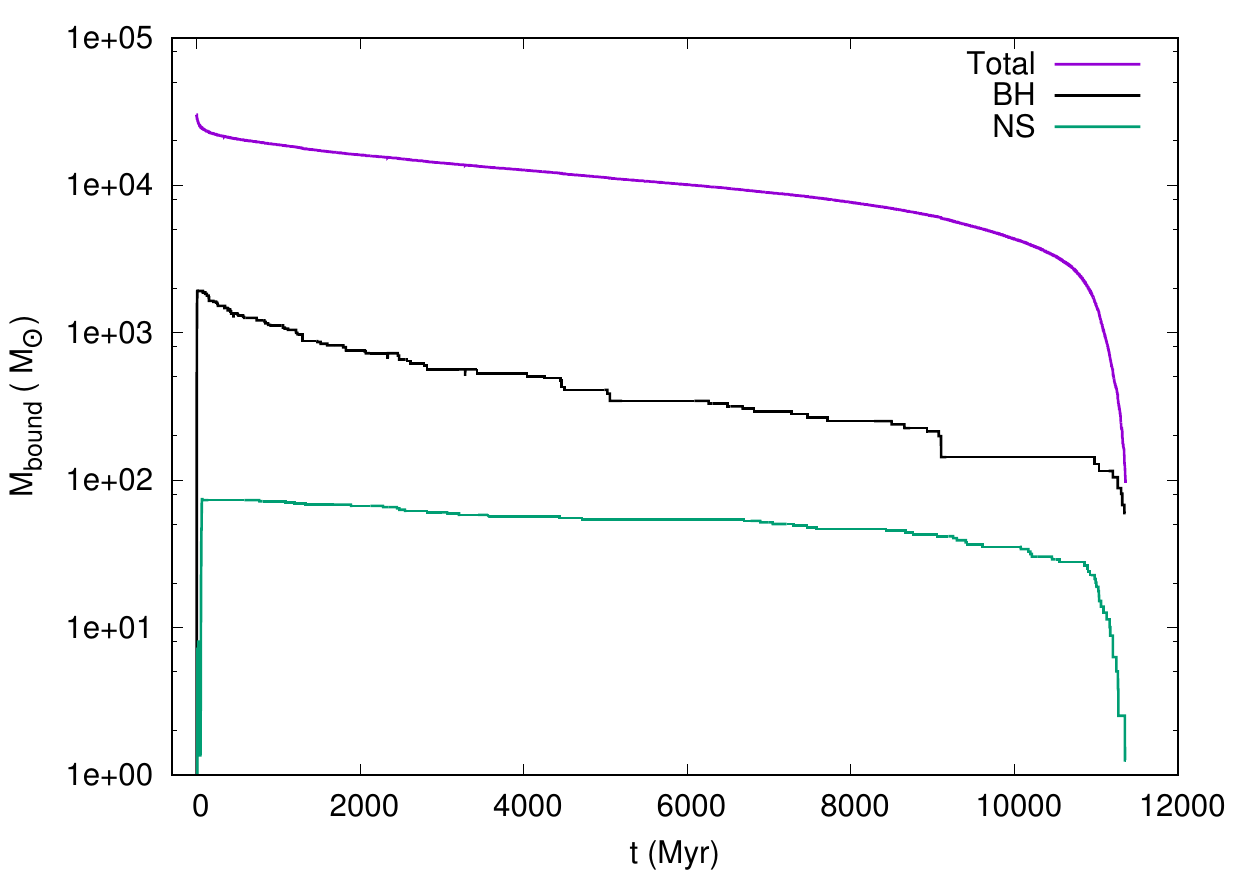}
\includegraphics[width=8.0cm,angle=0]{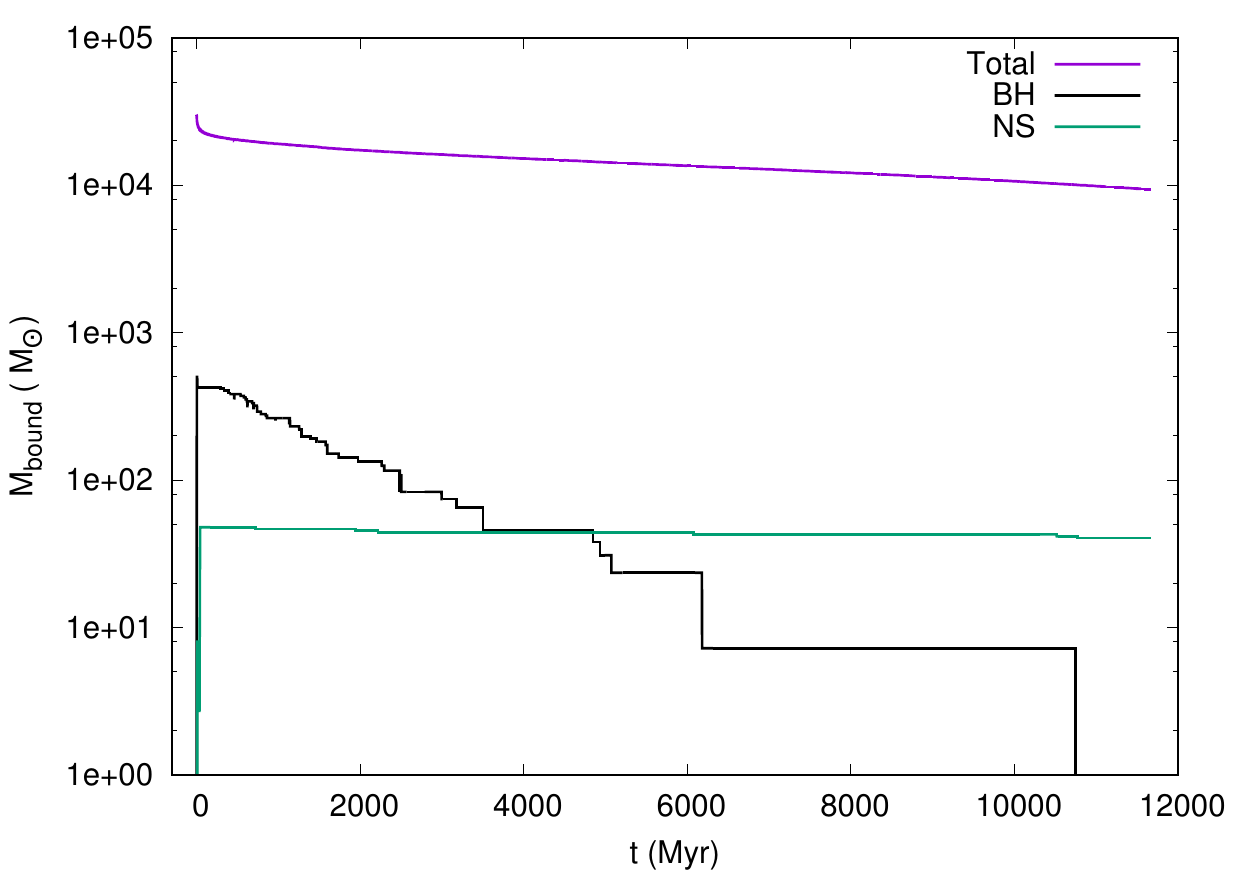}
\caption{Examples of computed evolution, for models with $\mcl(0)\approx3\times10^4\Ms$ and
	$\rh(0)\approx2$ pc having metallicities $Z=0.05\Zs$ (left column) and $Z=\Zs$ (right column);
	see Table~\ref{tab:comp}.
	{\bf Top panels:} time evolutions of the 1\%, 2\%, 5\%, 10\%, 20\%, 30\%, 40\%, 50\%, 62.5\%, 75\% and 90\%
	Lagrange radii ($R_f$ denotes the $(f\times100)$\% radius),
	{\bf Middle panels:} evolutions of the numbers of BHs and NSs bound to the
	cluster, {\bf Bottom panels:} evolutions of the total bound cluster mass (blue line) and the total masses
	of the BHs and NSs bound the cluster (black and green lines respectively).}
\label{fig:lagrng_30k}
\end{figure*}

\begin{figure*}
\centering
\includegraphics[width=8.0cm,angle=0]{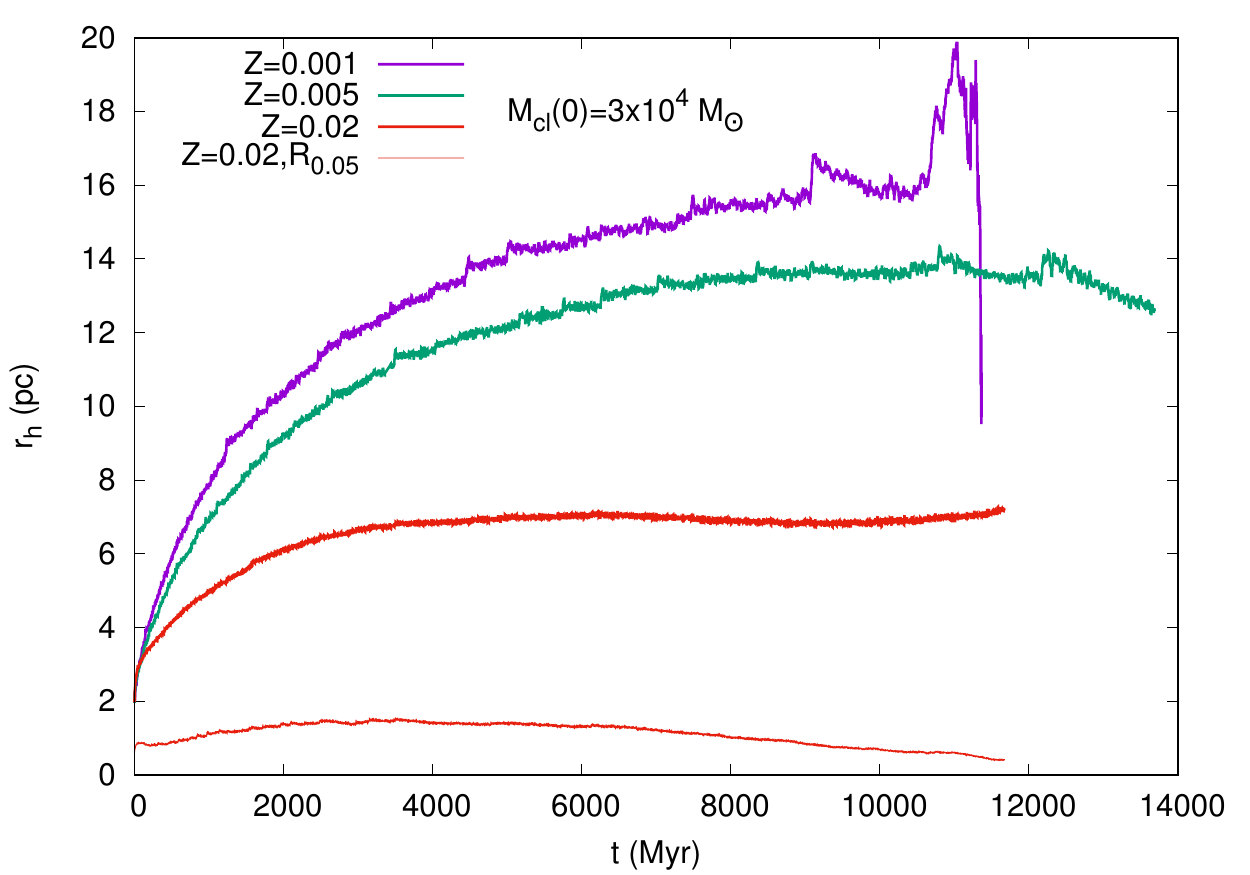}
\includegraphics[width=8.0cm,angle=0]{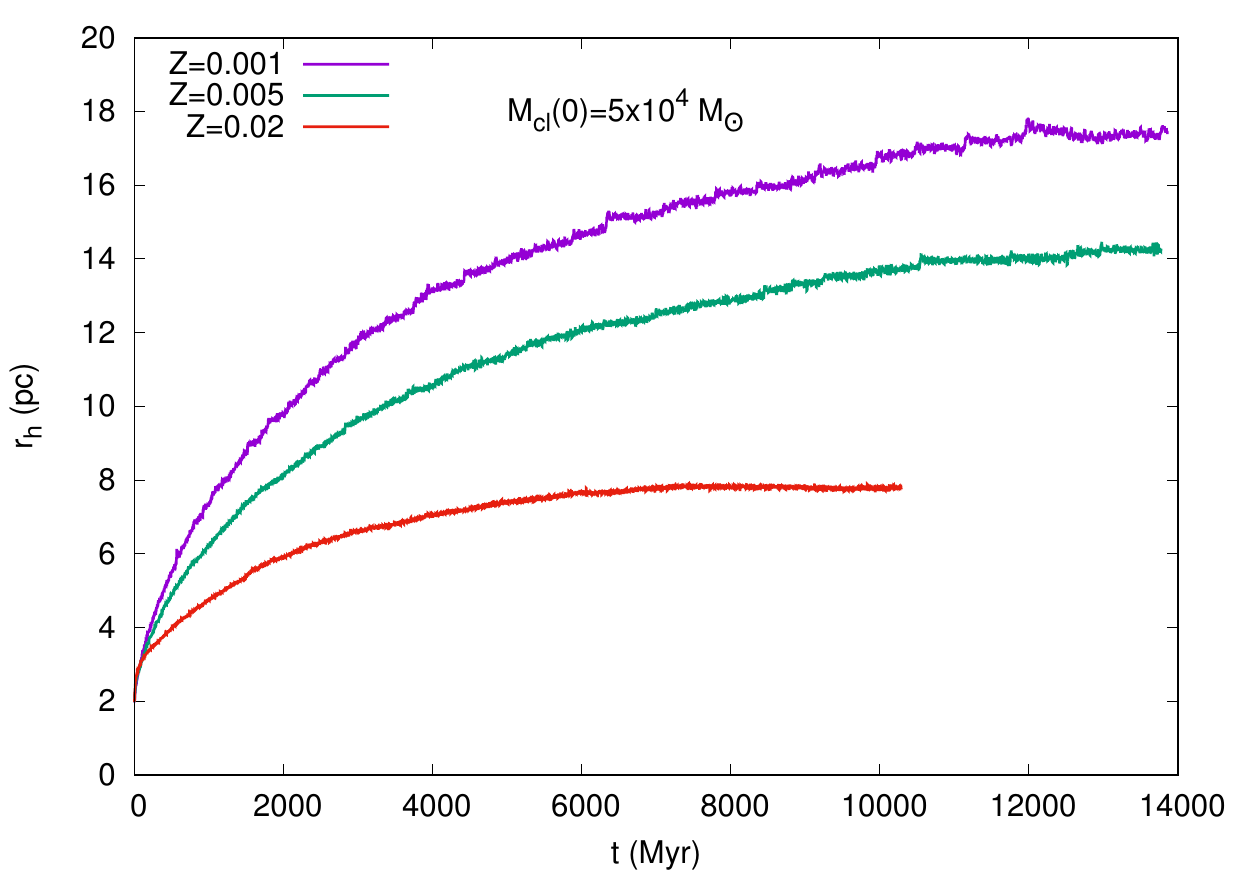}
\includegraphics[width=8.0cm,angle=0]{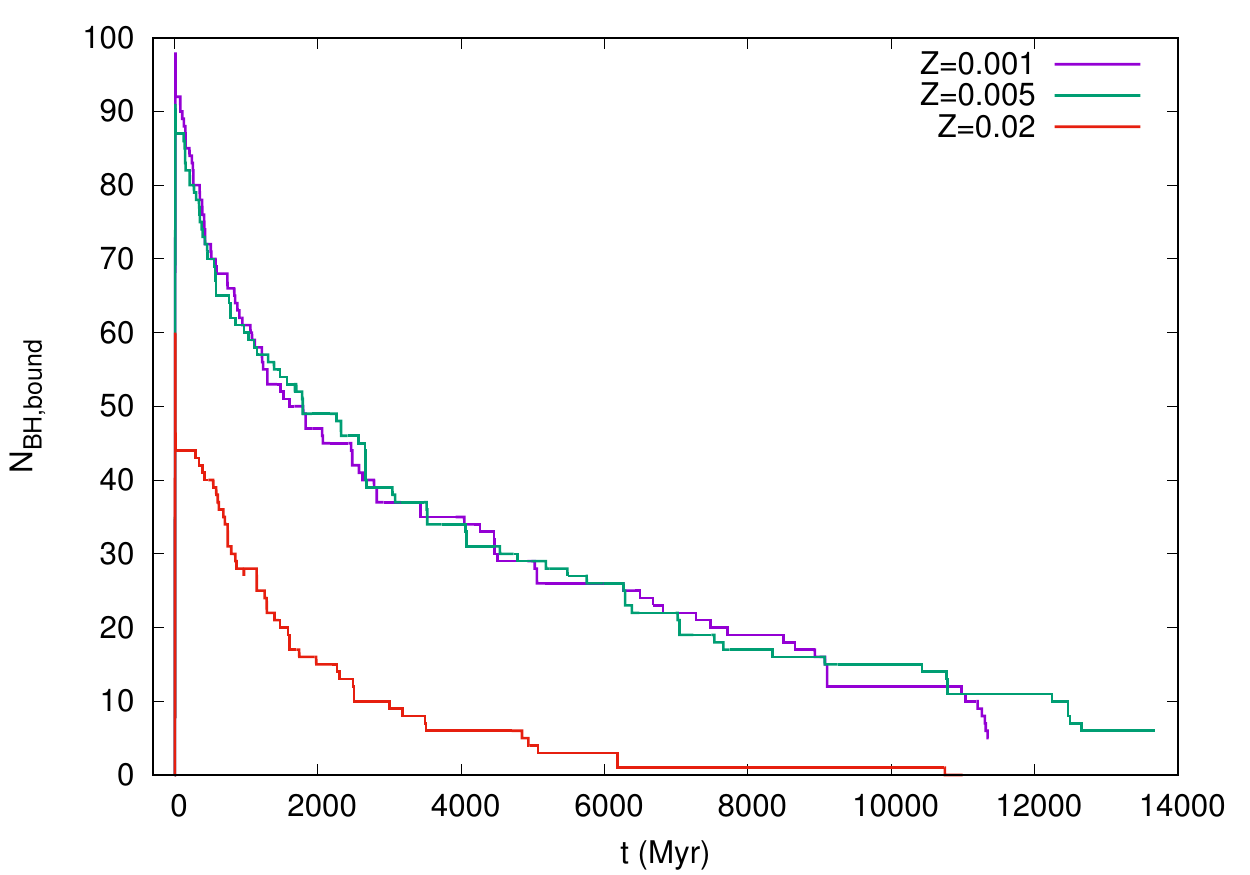}
\includegraphics[width=8.0cm,angle=0]{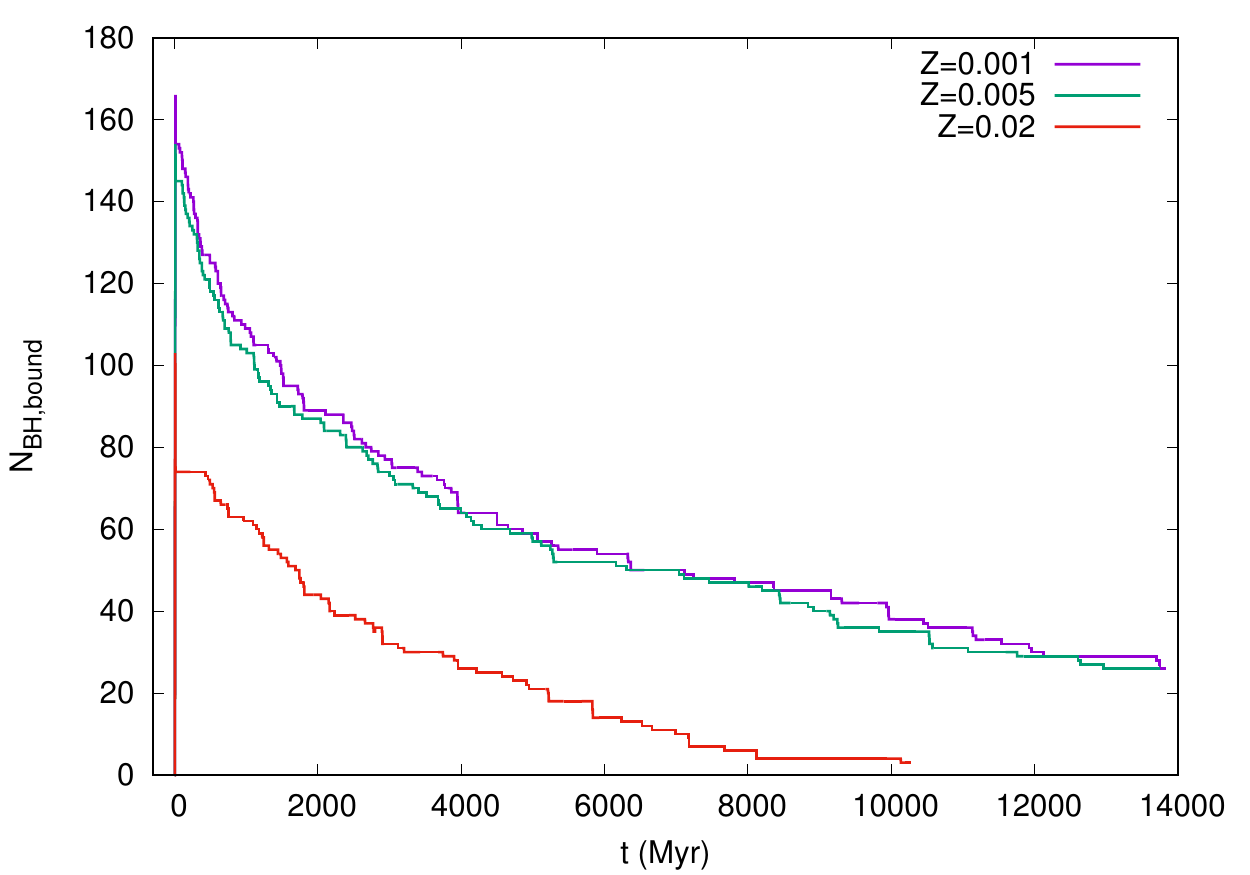}
\caption{The time evolution of the half-mass radius (50\% Lagrange radius $R_{0.5}$), $\rh$ (top row),
and the number of BHs bound to the cluster, $\nbhbound$ (bottom row), as a function of metallicity,
Z, for computed model clusters with $\mcl(0)\approx3\times10^4\Ms$ (left column)
and $\mcl(0)\approx5\times10^4\Ms$ (right column),
each having $\rh(0)\approx2$ pc (see Table~\ref{tab:comp}). The 5\% Lagrange radius, $R_{0.05}$,
is also plotted for the $\mcl(0)\approx3\times10^4\Ms$, $Z=\Zs$ model (top left panel)
to indicate that it has just arrived at its (second) core-collapsed phase.}
\label{fig:hmr_bhnum_Z}
\end{figure*}

\section{Model calculations}\label{calc}

\subsection{The \nbseven N-body evolution program}\label{nbprog}

The \nbseven code is an immediate descendant of the widely-used
\nbsix direct N-body evolution code
\citep{2003gnbs.book.....A,Nitadori_2012}. \nbseven utilizes
the Algorithmic Regularization Chain (ARC) of \citet{Mikkola_1999} instead of
the the classic Chain Regularization in {\nbsix} \citep{Mikkola_1993,2003gnbs.book.....A}.
This enables a more thorough and reliable treatment of multiple systems that continue
to form dynamically in any dense environment and influence the dynamics,
especially of those involving one or more massive objects like BHs.
\nbseven otherwise follows similar numerical strategies like \nbsix, namely,
a fourth-order Hermite integrator is used to accurately advance the trajectories
of each star subjected to the resultant force from rest of the bodies.
To ease the consequent $\propto N^3$ dependence in computing time, a
neighbour-based scheme is utilized for computing the force contributions
\citep{Nitadori_2012} at the shortest time intervals (the ``irregular'' force/steps).
At longer time intervals (the ``regular'' force/steps), all members in the
system are included for the force evaluation. The inexpensive (but numerous) irregular forces
are computed using parallel processing in regular single-node workstation CPUs, while the much more
expensive regular force
calculations are done on {\tt CUDA}\footnote{Compute-Unified Device Architecture}-enabled high-performance
GPUs\footnote{All computations in this work are done on
workstations equipped with quad-core {\tt AMD} processors and {\tt NVIDIA}'s
{\tt Fermi} and {\tt Kepler} series GPUs.}. The diverging gravitational forces
during close passages and in binaries are dealt with two-body or KS regularization
\citep{2003gnbs.book.....A} and higher-order multiples are treated with the ARC.

Like its predecessors, \nbseven utilizes the semi-analytic stellar evolution code
{\bse} \citep{Hurley_2000,Hurley_2002}\footnote{\bse is the binary stellar
evolution code and its counterpart for evolving single stars is called
\sse, which are available as separate standalone packages. The same wind mass loss
and remnant formation recipes (see Sec.~\ref{newwind}) can be applied
to both. In the N-body code, the evolutionary subroutines of \sse and \bse
are integrated into the various \nbseven subroutines and we will
simply denote the stellar-evolutionary part of \nbseven as \bse.}
to evolve each star and form their remnants.
The stellar parameters or each
star (including any tidal effect if it is a member of a binary or a multiple) are
updated simultaneously with its trajectory integration. That way the effects of
stellar evolutionary mass loss, via winds and supernovae, are naturally incorporated
in a calculation.

\begin{figure*}
\centering
\includegraphics[width=8.0cm,angle=0]{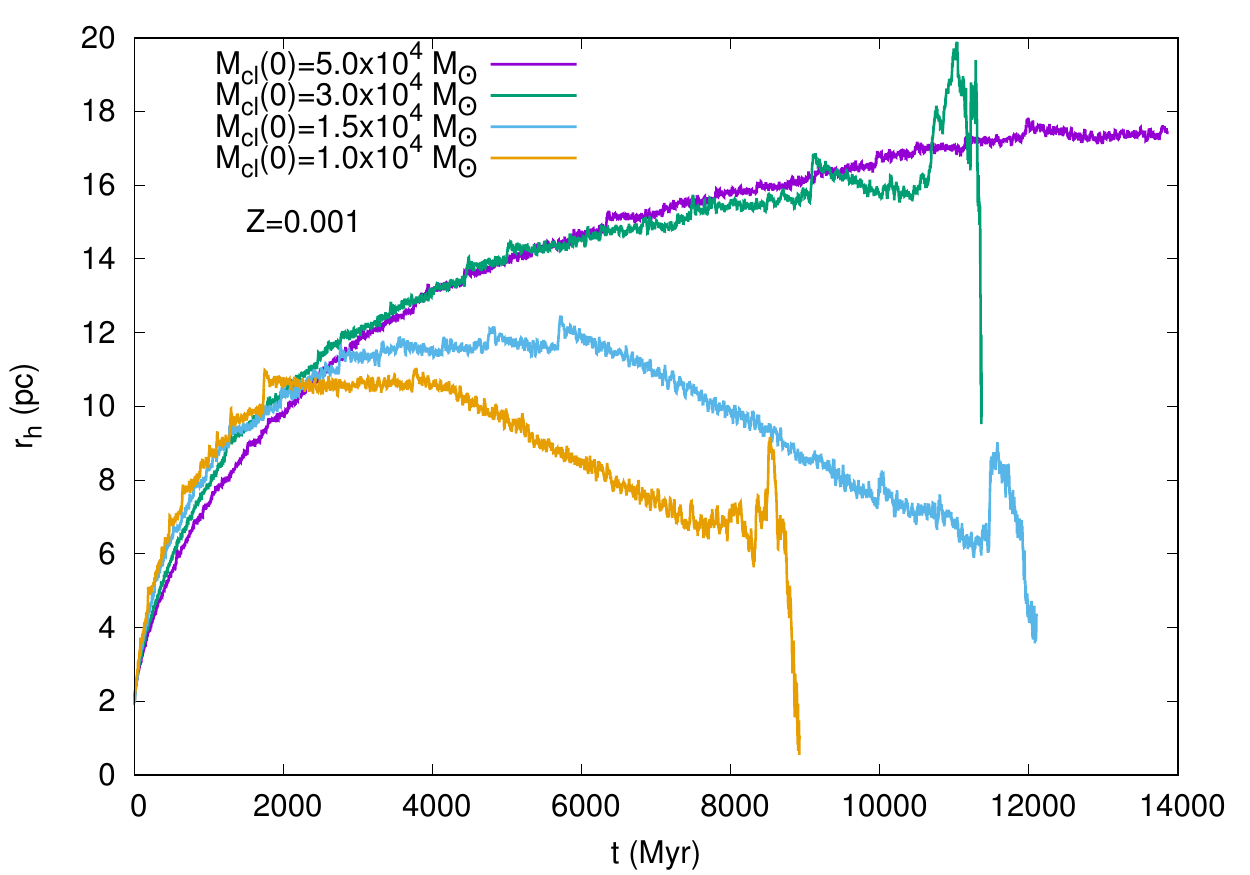}
\includegraphics[width=8.0cm,angle=0]{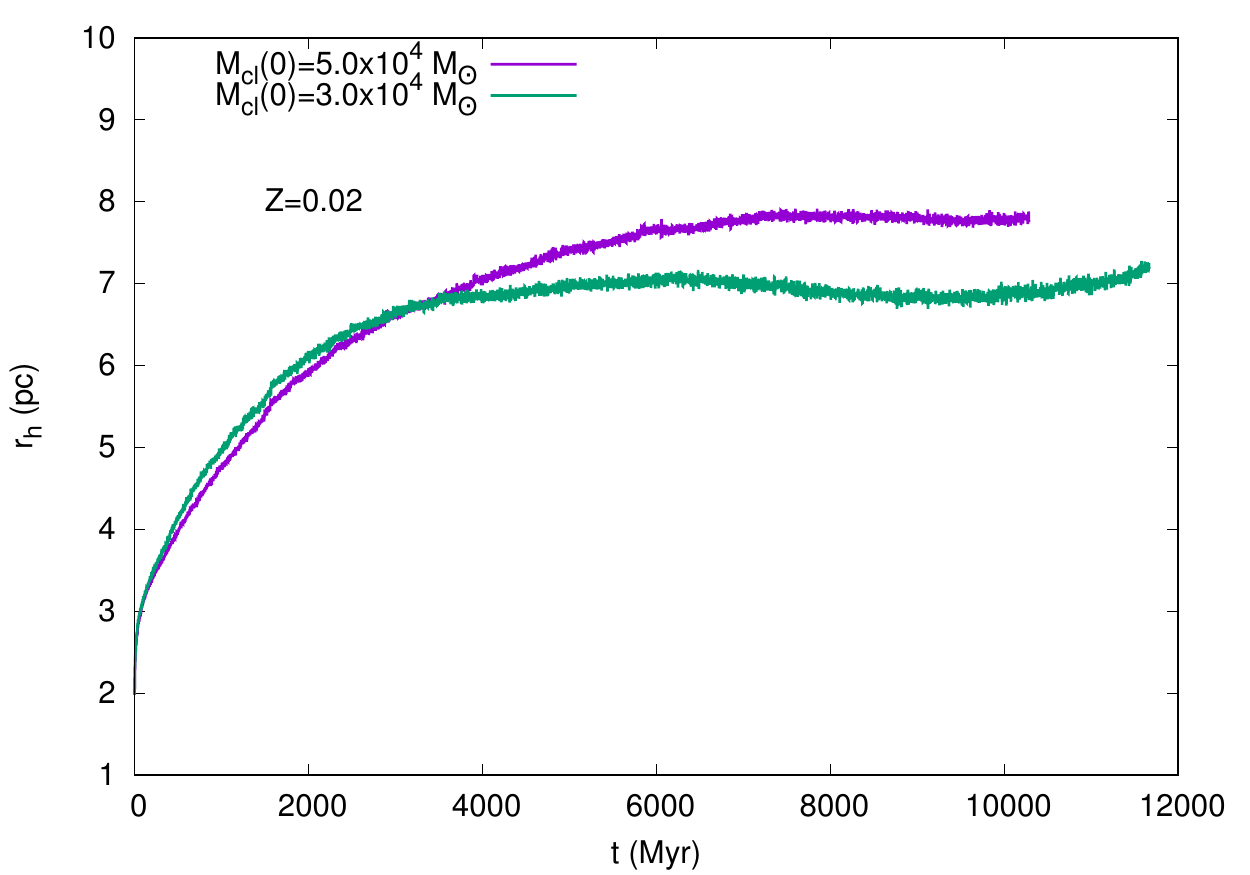}
\includegraphics[width=8.0cm,angle=0]{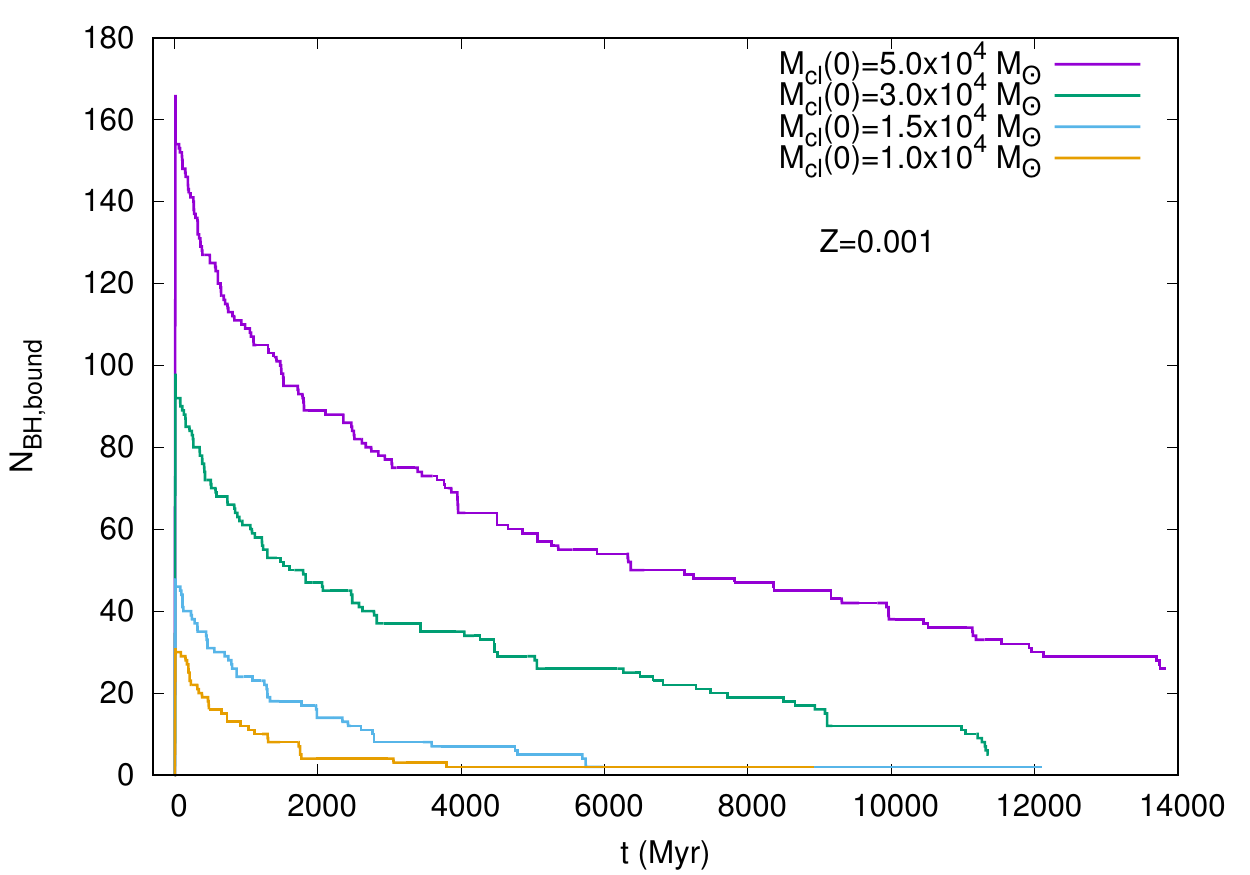}
\includegraphics[width=8.0cm,angle=0]{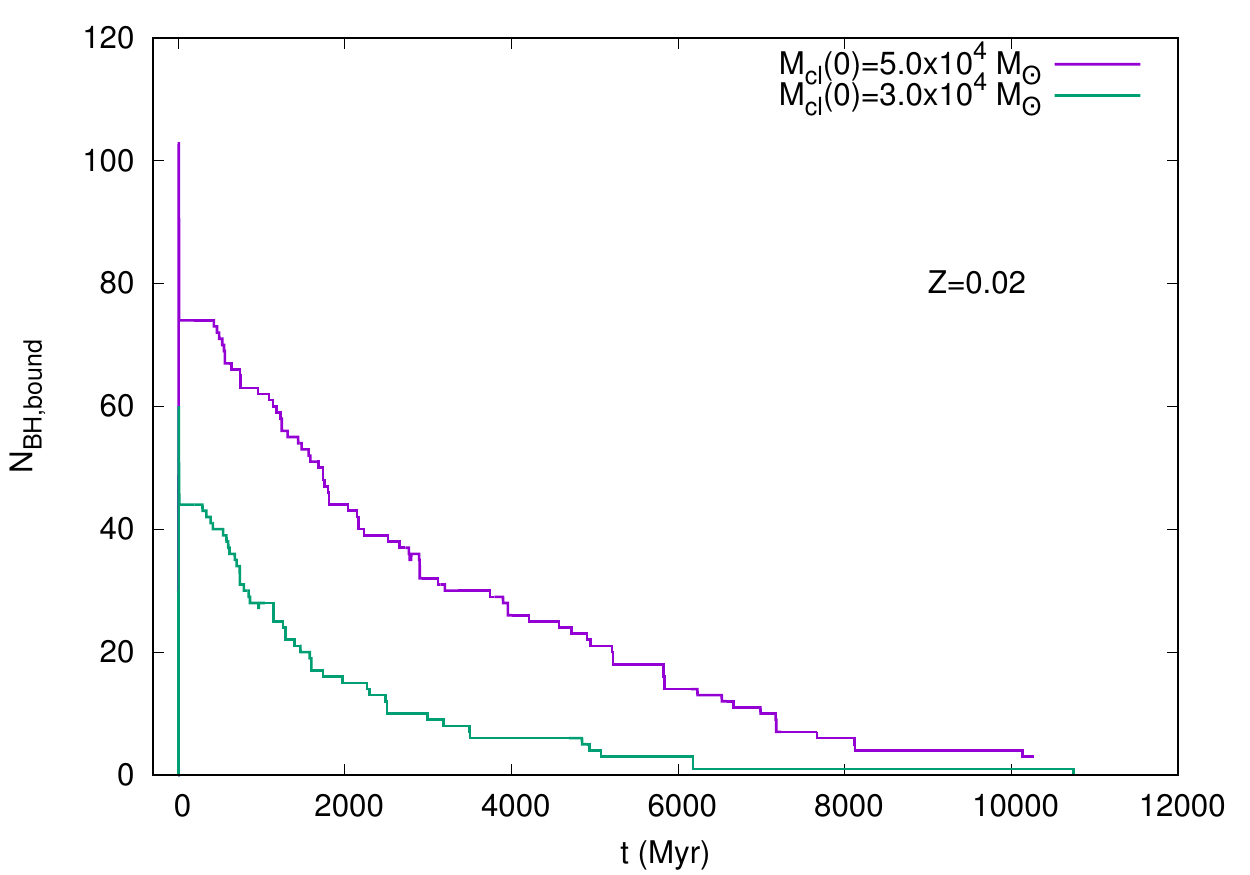}
\caption{The evolution of $\rh$ (top row) and $\nbhbound$ (bottom row),
	as a function of $\mcl(0)$, for computed models with $Z=0.05\Zs$ (left column)
	and $Z=\Zs$ (right column), each having $\rh(0)\approx2$ pc (see Table~\ref{tab:comp}).}
\label{fig:hmr_bhnum_M}
\end{figure*}

An important aspect of \nbseven is its general relativistic (GR) treatment, when an NS or/and
a BH is a member of binary or a multiplet. The relativistic treatment, following
the Post-Newtonian (PN) approach, is included in the ARC procedure \citep{Mikkola_2008}.
In principle, PN-1.5 (GR periastron precession), PN-2.5 (orbital shrinking due to
GW radiation) and PN-3.5 (spin-orbit coupling) order terms are included in the ARC; see
\citet{2012MNRAS.422..841A} for the key elements of the implementation in \nbseven
and \citet{Brem_2013} for an alternative approach (in {\tt NBODY6++}). This allows for
on-the-fly GR orbital modifications and coalescences of relativistic subsystems (typically
a binary or a triple containing one or more BH/NS) that are bound to the system.
The latest implementation generally shows reasonable energy check
(typical relative energy change
$\sim10^{-6}$) even during extreme relativistic events such as a BBH coalescence within a triple.
In the present
computations (see Sec.~\ref{runs}), however, the PN terms up to order 2.5 are applied, as activating the spin terms
would make these computations, which typically contain one or more relativistic subsystems nearly
all of the time, much more slower\footnote{In addition to the PN treatment in the ARC procedure,
GR treatment is also applied perturbatively in case of weakly relativistic subsystems
(a wide but eccentric BBH or a hierarchical BH-triple). This procedure is still
under development and shows unstable behaviour when
a large number of BHs are present (Sverre Aarseth, private communication), as in
the present models, which is why the perturbed-PN has also not been applied here.}.
The spin terms would have modified the times of the BBH coalescences occurring within the cluster
(see Secs.~\ref{intro} \& \ref{bhbin}) to some extent, however, this is
not critical in the present context due to the statistical nature of the dynamically-induced
BBH coalescences.

In reality, a BBH would typically receive a substantial GW merger kick during its
inspiral phase ($\sim100-1000\kmps$; \citealt{Campanelli_2007,Hughes_2009}),
due to the presence of the BHs' spins. 
This would cause the newly-formed merged BH to escape from the cluster almost inevitably,
and it would hardly have a chance to participate in dynamical encounters further.
This situation is mimicked by applying a velocity kick onto the merged BH, immediately after a
coalescence happens within the cluster (Sec.~\ref{bhbin}). To avoid large energy error, the applied kick
is kept only marginally above the escape speed; $\approx5$ times the central RMS speed.
This is still enough the eject the merged BH out of the cluster, in $\sim10$ dynamical
times; in reality a BBH coalescence product would typically escape at much higher
speed\footnote{In test calculations, it is found that even if the merged product is retained,
it does not necessarily take part in further relativistic coalescences.}.

\begin{figure*}
\centering
\includegraphics[width=8.0cm,angle=0]{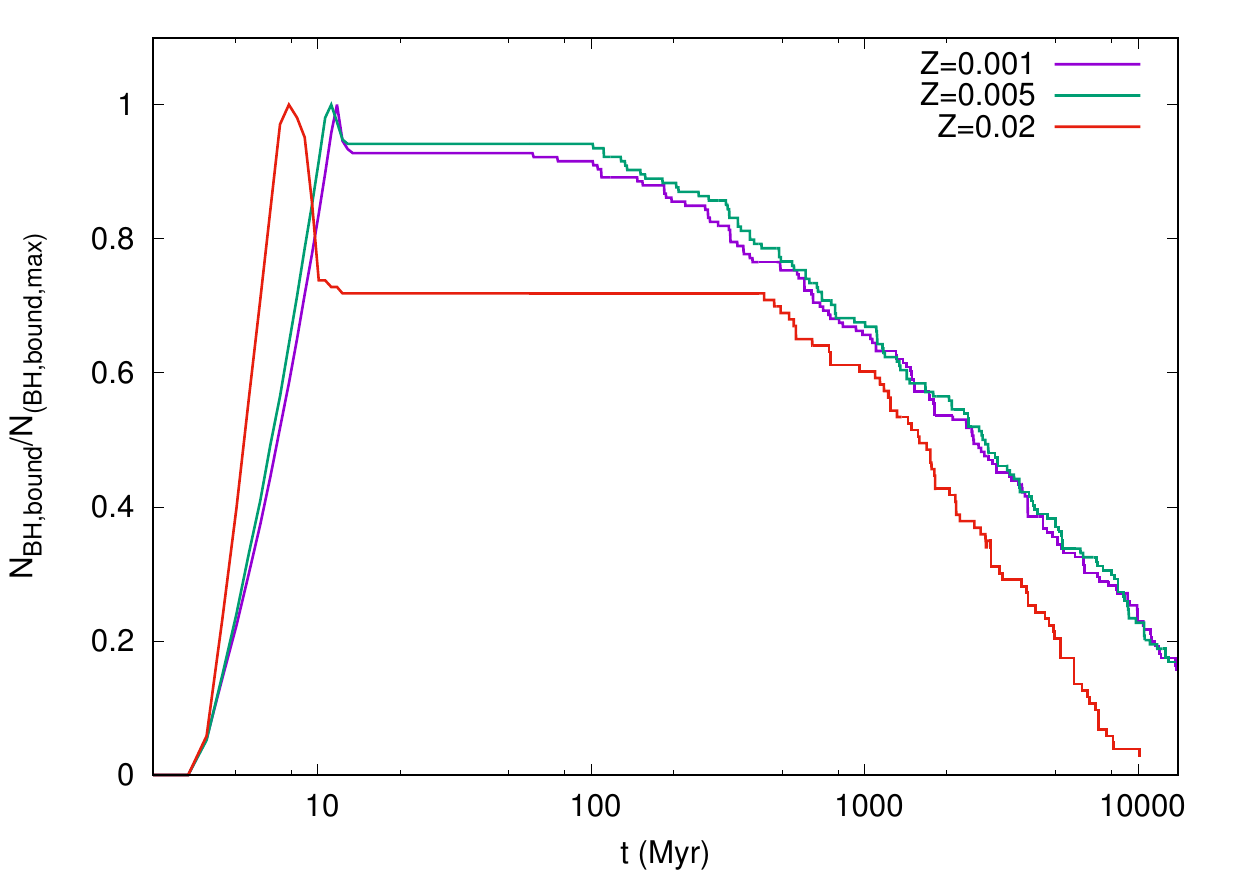}
\includegraphics[width=8.0cm,angle=0]{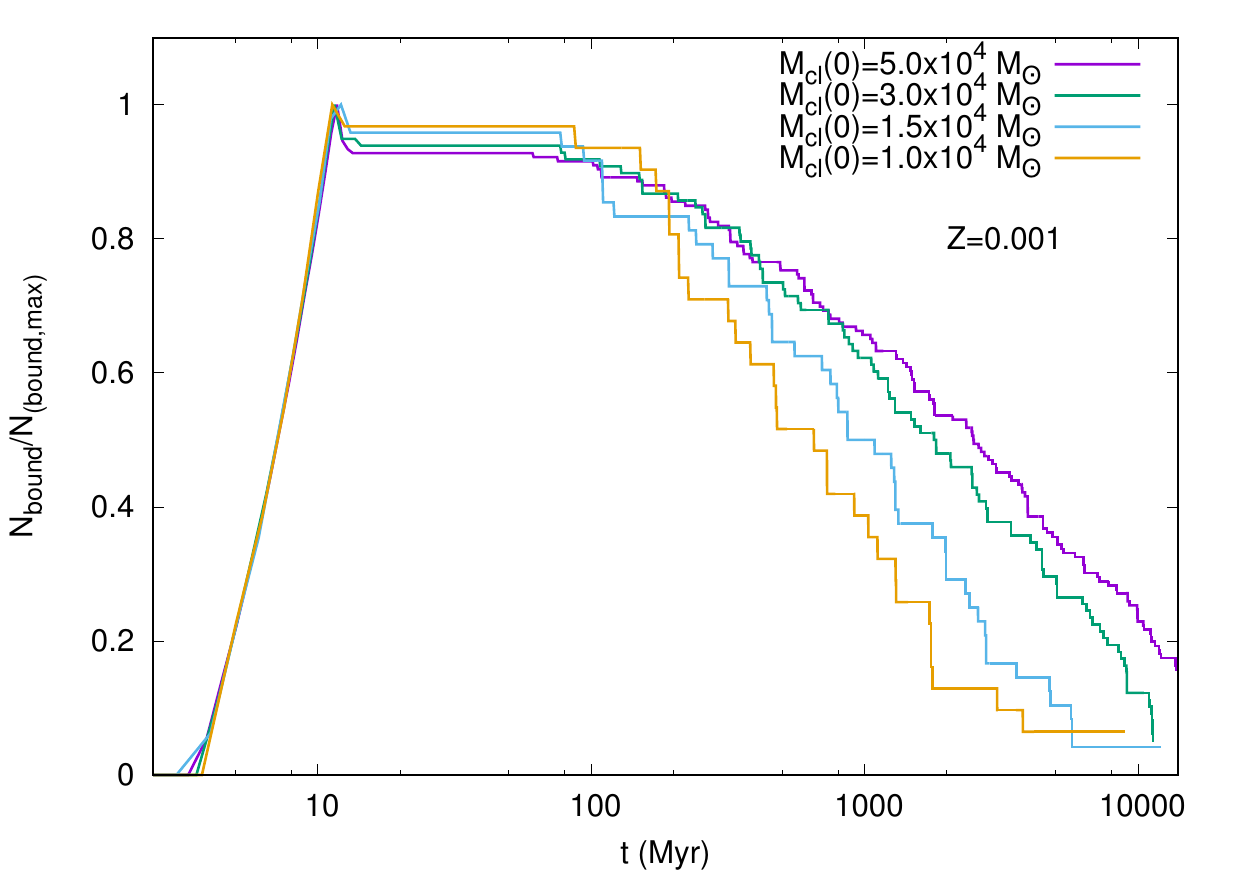}
	\caption{{\bf Left:} The time evolution of the fraction of BHs bound to the cluster
	as a function of $Z$, for the computed models with $\mcl(0)\approx5\times10^4\Ms$,
	$\rh(0)\approx2$ pc. {\bf Right:} The time evolution of the fraction of BHs bound to the
	cluster as a function of $\mcl(0)$, for computed models with $Z=0.05\Zs$ and
	$\rh(0)\approx2$ pc (see Table~\ref{tab:comp}).}
\label{fig:bhfrac}
\end{figure*}

\begin{table*}
\centering
\caption{Summary of the model stellar clusters in this work, whose evolutions are
	computed using \nbseven. The columns from
	left to right respectively denote: (a) initial mass, $\mcl(0)$, of the model cluster,
	(b) initial half-mass radius, $\rh(0)$, (c) metallicity, $Z$, (d) number of 
	(triple-mediated) binary black hole (BBH) coalescences, $\nmrgin$, that occurred within the clusters,
	(e) number of BBH coalescences (in BBHs ejected from the clusters), $\nmrgout$, that occurred
	outside the clusters within the Hubble time. For the BBHs that have
	undergone coalescence, the masses of the corresponding binary members are
	indicated in parentheses in the columns (d) and (e).
	}
\label{tab:comp}
\begin{tabular}{lccll}
	\hline
	\hline
	\mcl(0)/\Ms     & \rh(0)/pc & $Z/\Zs$ & \nmrgin                   & \nmrgout\\
	\hline
	$5.0\times10^4$ & 2.0       & 0.05    & 1 ($24.3\Ms+17.7\Ms$)     & 1 ($26.0\Ms+42.8\Ms$)\\
	\hline
	$5.0\times10^4$ & 2.0       & 0.25    & 1 ($34.5\Ms+22.7\Ms$)     & 0                    \\
	\hline
	$5.0\times10^4$ & 2.0       & 1.00    & 3 ($9.0\Ms+7.5\Ms$)       & 0                    \\
		        &           &         &   ~~~($10.6\Ms+9.4\Ms$)   &                      \\
		        &           &         &   ~~~($9.1\Ms+9.0\Ms$)    &                      \\
	\hline
	\hline
	$3.0\times10^4$ & 2.0       & 0.05    & 1 ($38.1\Ms+25.9\Ms$)     & 2 ($25.7\Ms+13.8\Ms$) \\
	                &           &         &                           & ~~~($23.6\Ms+22.3\Ms$) \\
	\hline
	$3.0\times10^4$ & 2.0       & 0.25    & 0                         & 2 ($35.2\Ms+20.3\Ms$) \\
	                &           &         &                           & ~~~($15.7\Ms+12.2\Ms$) \\
	\hline
	$3.0\times10^4$ & 2.0       &  1.00   & 1 ($10.6\Ms+9.0\Ms$)      & 0                       \\
	\hline
	\hline
	$1.5\times10^4$ & 2.0       & 0.05    & 1 ($49.4\Ms+30.9\Ms$)     & 0                      \\
	$1.5\times10^4$ & 1.0       & 0.25    & 0                         & 0                      \\
	\hline
	\hline
	$1.0\times10^4$ & 2.0       & 0.05    & 0                         & 0                      \\
	$1.0\times10^4$ & 1.0       & 0.05    & 1 ($43.6\Ms+34.5\Ms$)     & 0                      \\
	$1.0\times10^4$ & 1.0       & 0.25    & 0                         & 0                      \\
	\hline
	\hline
	$0.7\times10^4$ & 1.0       & 0.05    & 0                         & 0                      \\
	\hline
	\hline
\end{tabular}
\end{table*}

\subsection{The new wind prescription:
remnant masses and natal kicks}\label{newwind}

The masses of the BHs, and how many of them receive low natal kicks so that
they can retain in the parent cluster at their birth, determines the effectiveness
of the BH-engine (Sec.~\ref{intro}).
For a given isolated progenitor star, the remnant BH mass is determined
by the entire history of the wind mass loss until the pre-core-collapse stage and
also on the material ``fallback'' onto the remnant during the supernova.
The BH's natal kick will be diminished by the amount of fallback; in particular
there will be no natal kick if the fallback is 100\%, \ie, the entire
pre-core-collapse star implodes into a BH (a ``failed supernova''; there
might still be a small kick due to the escape of neutrinos).
If the progenitor is initially in a close binary,
its mass loss (or gain) and hence the BH mass can additionally
be influenced by any mass transfer or by tidal heating effect from its companion.
Unfortunately, to date, massive-stellar winds, the mechanisms of
core-collapse supernovae and material fallback are still poorly understood
or constrained.

In this work, we will adopt the semi-analytic remnant formation and wind
prescriptions of \citet{Belczynski_2008,Belczynski_2010}.
These rather widely used prescriptions are based on empirically-determined
wind mass loss formulae of \citet{Vink_2001} for O/B-stars,
metallicity-dependent winds (including suppression due to clumping)
of \citet{Vink_2005} for Wolf-Rayet (naked helium) stars
and metallicity-independent (suppressed) wind for Luminous Blue Variable stars;
see the formulae (6)-(9) and their explanations in \citet{Belczynski_2010}
(hereafter B10). After the wind mass loss until the formation of the pre-supernova
core, the remnant (NS or BH) mass is determined based on the {\tt CO} and 
{\tt FeNi} core mass and the amount of fallback onto it as in
\citet{Belczynski_2008}; see their Eqns.~(1) \& (2).

Such partly empirical and partly physically-motivated wind and remnant formation
model combinations constitute perhaps the
only known way to plausibly obtain $\approx30\Ms$
BHs as observed in the GW150914 event, at $\approx\Zs/4$ metallicity that
is plausible for the Local Universe (redshift $z<0.2$) where this event must have
occurred \citep{2016PhRvL.116f1102A,2016ApJ...818L..22A}. Overall, the
B10 winds are much weaker at any metallicity than the standard
\citet{Hurley_2000} (hereafter H2K) winds that is adopted by default in the
{\bse} (see Sec.~\ref{nbprog}) routine, the latter yielding up
to $\lesssim25\Ms$ BHs at metallicity as low as $\Zs/100$,
that is unlikely to occur in the local universe. That way
the B10 wind prescription more plausibly yields the mass range
of BHs inferred from the three LIGO events, than the H2K prescription
(for the same \citealt{Belczynski_2008} remnant-formation prescription),
which is why the former is preferred in this study.

The birth kicks of the remnants are assigned based on their type, mass and
the amount of material fallback, as derived from the above prescriptions.
The NSs produced from core-collapse supernovae
are given large kicks of $\approx265\kmps$, as inferred from observations of radio
pulsars in the Galactic field \citep{Hansen_1997,Hobbs_2005}.
The BHs formed without or partial fallback are
then assigned diminished kicks based on their final masses, that are scaled
from the above NS kick assuming linear momentum conservation.
For pre-supernova {\tt CO} core mass of $M_{\tt CO}=7.6\Ms$,
the fallback is taken to become complete based on
the studies by \citet{Fryer_1999,Fryer_2001}\footnote{In the present prescription, the fallback
fraction grows linearly between $5\Ms \leq M_{\tt CO} \leq 7.6\Ms$.};
the entire (pre-supernova) star is assumed to collapse directly
into a BH for $M_{\tt CO}\geq7.6\Ms$ when zero natal kick is assigned.
NSs are also allowed to form via Electron Capture Supernovae
(ECS; \citealt{Podsiadlowski_2004}) as in \citet{Belczynski_2008}, which are of
$\approx1.26\Ms$ and are as well assigned zero natal kick. For the model
clusters considered in this study (with masses $\mcl(0)\leq5\times10^4\Ms$; see Table~\ref{tab:comp}),
only the direct collapse BHs and ECS NSs will retain in the cluster at their
birth. All core-collapse NSs and almost all other BHs will escape right after
their formation, due to their high natal kicks.

The above ``new wind'' and kick recipes have already been implemented in the
{\tt StarTrack} \citep{Belczynski_2002,Belczynski_2008}
semi-analytic stellar-evolutionary code
and are now implemented in the {\bse}, that is integrated with the \nbseven
(in the current version of \nbseven,
different elements of the above wind prescriptions are available as
options alternative to the default H2K recipe). As
pointed out above, there are uncertainties in nearly all physical factors
determining the remnant masses and their natal kicks. Therefore the
presently adopted schemes can at most be taken as a physically plausible
implementation of the present empirical and theoretical knowledge of
stellar winds and remnant formation, which is also widely utilized
(\eg, in \citealt{Morscher_2015,2016arXiv160300884C}).
Note that a similar wind and remnant formation schemes are adopted
by \citet{Spera_2015} but in a different stellar evolution and population synthesis
code, namely, the {\tt SEVN}; in this study, however, we will limit ourselves
to the \bse as it is adapted to the \nbseven.

Fig.~\ref{fig:bhmass} (top panel) shows the remnant mass as a function of
zero-age main sequence (ZAMS) mass at different metallicities, $Z$, as obtained
by the {\nbseven}-adapted {\bse} for the presently
assumed stellar wind and remnant formation recipes (thin lines).
They agree reasonably well with those of B10 for the same values of $Z$ (thick lines),
which have been obtained using {\tt StarTrack}.
Fig.~\ref{fig:bhmass} (bottom panel) shows the mass distributions
of the BHs that remain bound to the $\mcl(0)=5\times10^4\Ms$ model clusters
computed here (see Table~\ref{tab:comp}) after their formation, \ie, receive zero or
low natal kicks (these mass distributions are obtained shortly after
the last BH formed at $\approx10$ Myr, when most of the bound BHs are still
segregating towrads the cluster's center but the unbound ones
have already escaped through the tidal radius, and when the dynamical ejections of the
BHs are yet to start; see below). As expected from the ZAMS mass-BH mass
relations, lower $Z$ would result in a wider BH mass spectrum. In all cases,
only the BHs from the first mass bin are depleted due to
natal kicks, the rest receive zero kicks; hence, the lower the $Z$ is,
the higher is the initial BH retention, in the present scheme.

\subsection{Model computations}\label{runs}

\begin{figure*}
\centering
\includegraphics[width=12.0cm,angle=0]{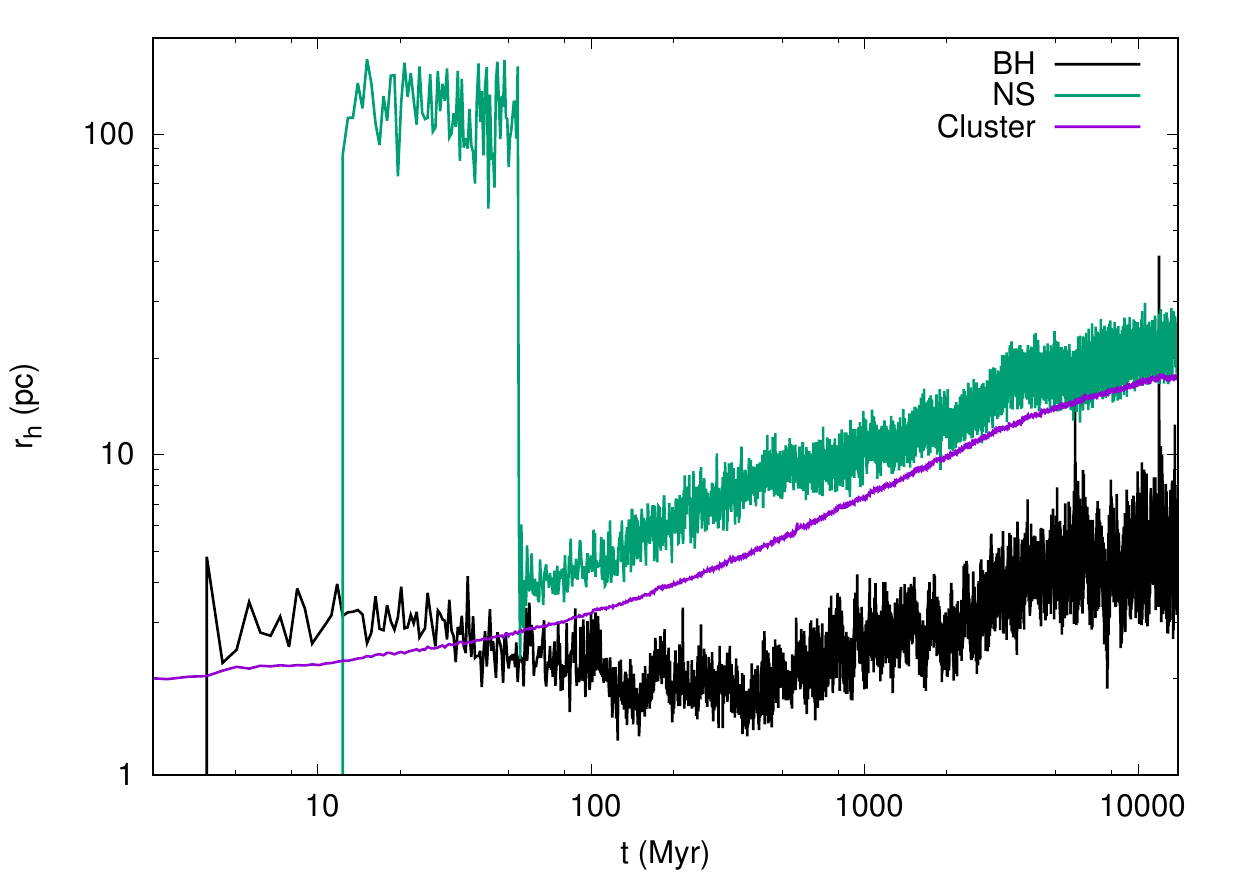}
	\caption{The time evolution of the half-mass radii of the whole cluster (blue line),
	the bound BHs (black line) and NSs (green line) for the computed model with
	$\mcl(0)\approx5\times10^4\Ms$, $\rh(0)\approx2$ pc, $Z=0.05\Zs$.}
\label{fig:hmr_rem_50kZlow}
\end{figure*}

\begin{figure*}
\centering
\includegraphics[width=8.0cm,angle=0]{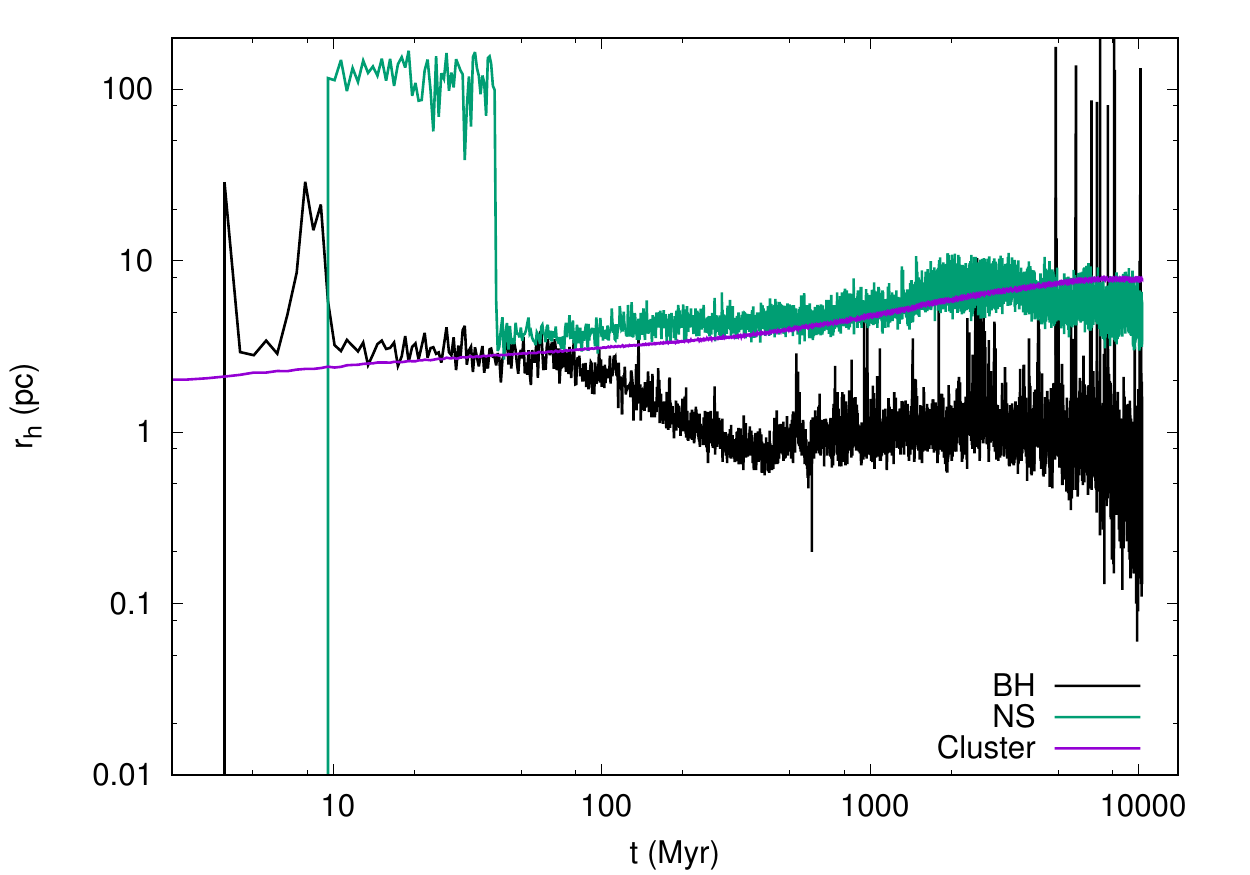}
\includegraphics[width=8.0cm,angle=0]{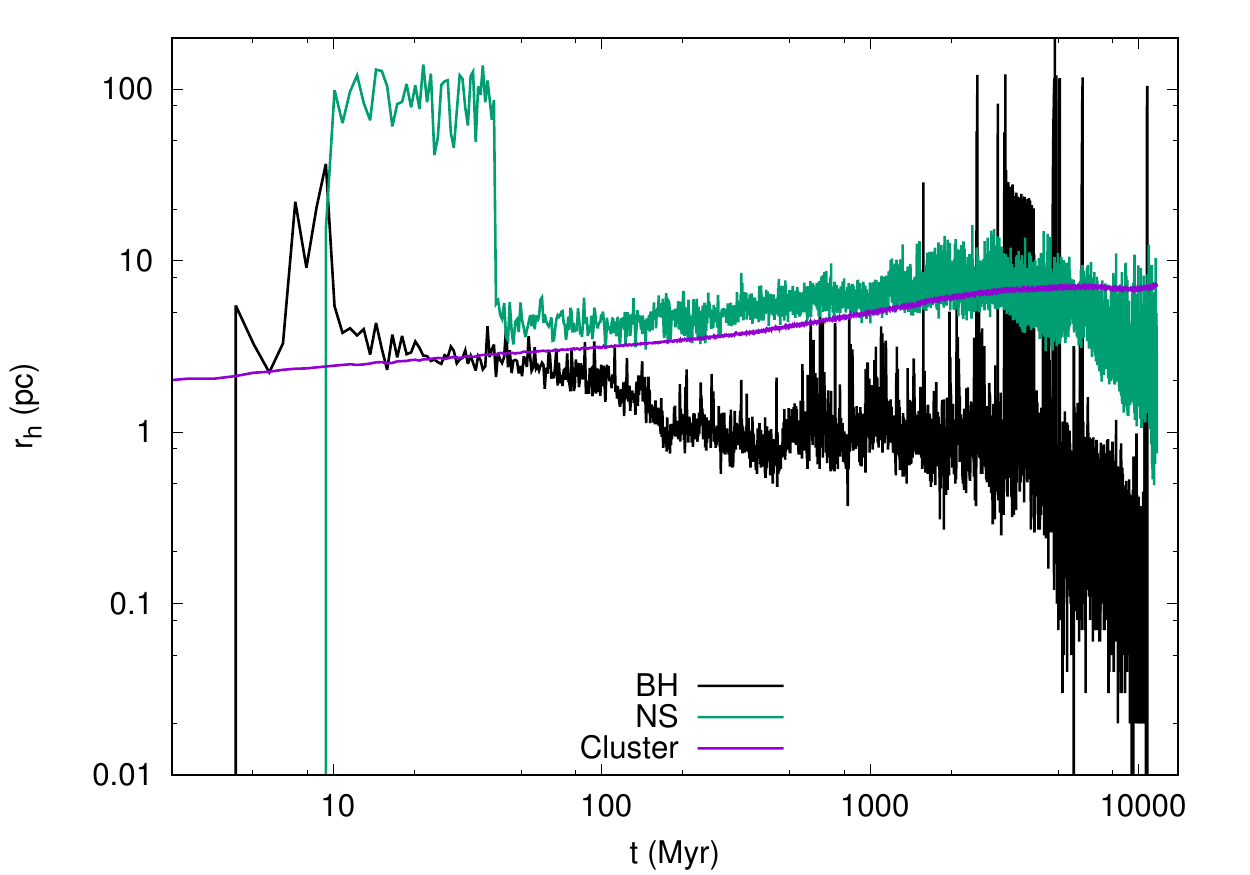}
	\caption{The same as in Fig.~\ref{fig:hmr_rem_50kZlow} but with $Z=\Zs$, for
	$\mcl(0)\approx5\times10^4\Ms$ (left) and $\mcl(0)\approx3\times10^4\Ms$
	(right). $\rh(0)\approx2$ pc for each.}
\label{fig:hmr_rem_Zsolar}
\end{figure*}

Table~\ref{tab:comp} lists the model computations for this study, which are
done using \nbseven. All the computed models are initially Plummer clusters with
masses $\mcl(0)\leq 5\times10^4\Ms$ and half-mass radius $\rh(0)=2$ pc; $\rh(0)=1$ pc
are also used for a few lower mass models. Such masses and sizes are typical
for Galactic and Local Group YMCs \citep{Banerjee_2016,PortegiesZwart_2010,Ryon_2015};
considering the masses of GCs, they
represent intermediate-mass and open stellar clusters (see Sec.~\ref{intro}).
From the point of view of the study of BH dynamics and dynamical BBH coalescence,
such mass range is relatively unexplored (see Sec.~\ref{intro} and references
therein), which is alone a good reason to study them here. However, reaching
higher initial masses would still have been interesting (see Sec.~\ref{bhbin}), and
not doing so here is mainly due to the computational costs involved
(most models are initiated with $\rh(0)=2$ pc instead of 1 pc for the same reason).
The initial mass function (IMF) of all the models are taken to be the canonical
one, satisfying the {\it observed} maximum stellar mass-cluster mass relation \citep{Weidner_2004,Kroupa_2013}. 
The metallicities of the models are varied between $0.05\Zs \leq Z \leq \Zs$;
$\approx0.05\Zs$ being the lowest metallicity {\it observed} in the Local Universe
\citep{SanchezAlmedia_2015,Rubio_2015}.

All models are evolved for 10.0-13.7 Gyr or until they dissolve completely,
beginning from their zero age. A solar neighbourhood-like, static external tidal field is
applied for each model. Note that this external field is just a
representative in this study, whose job is simply to remove any gravitationally unbound object from the
cluster's membership, and its use is of course a simplification. In reality, a cluster would have
gone through large changes in its environment over such long evolutionary times and hence
in the tidal field it is subjected to (see, \eg, \citealt{Renaud_2015b});
even our immediate cosmic neighborhood offers widely different environments to newly born
clusters, \eg, compare the external fields on to the YMCs of the Milky Way and of the
Magellanic Clouds. However, the growth of the host galaxy is
neighter fatal to the cluster nor alters the cluster's evolution drastically, as long
as it is adiabatic \citep{Renaud_2015a}. Over most of their evolutionary time, the computed models
underfill their tidal radii. 

Another simplification is the assumption of a monolithic, gas-free structure of the
model clusters right from their zero age, \ie, neglecting their assembly phase and
the effect of gas expulsion \citep{Longmore_2014,Banerjee_2015a}.
Depending on the duration of the assembly and
gas dispersal phase, which, in turn, would govern the mass segregation of the massive
stars and/or the stellar remnants, the dynamics of the BHs may or may not be
affected significantly by the assembly process. The current assumption of a
monolithic structure throughout is justified if the clusters undergo
``prompt assembly'' \citep{Banerjee_2015b}. 

Finally, all the initial models contain only single stars. This is also a simplification
done for the sake of computational ease. An appropriate initial binary fraction for O/B-stars would
be 50\%-70\%, to be consistent with what is observed in starburst clusters \citep{Sana_2011,Sana_2013},
which would be tedious to compute over such long evolutionary times. Recent Monte-Carlo calculations
(using 5-10\% O/B-stellar binary fraction; \citealt{2016arXiv160300884C})
infer that the properties of dynamically-produced
BBHs are nearly independent of the primordial binary content. This is because nearly all the BH
progenitor binaries widen significantly due to stellar wind and supernova mass loss, to either get
disrupted directly or get ionized by dynamical interactions with the surrounding stars, so
that the majority of the BHs become single anyway. Hence, the lack of primordial binaries
in our computed models would not pose a serious limitation as long as the dynamics of the
BHs are concerned.
 
\begin{figure}
\centering
\includegraphics[width=8.0cm,angle=0]{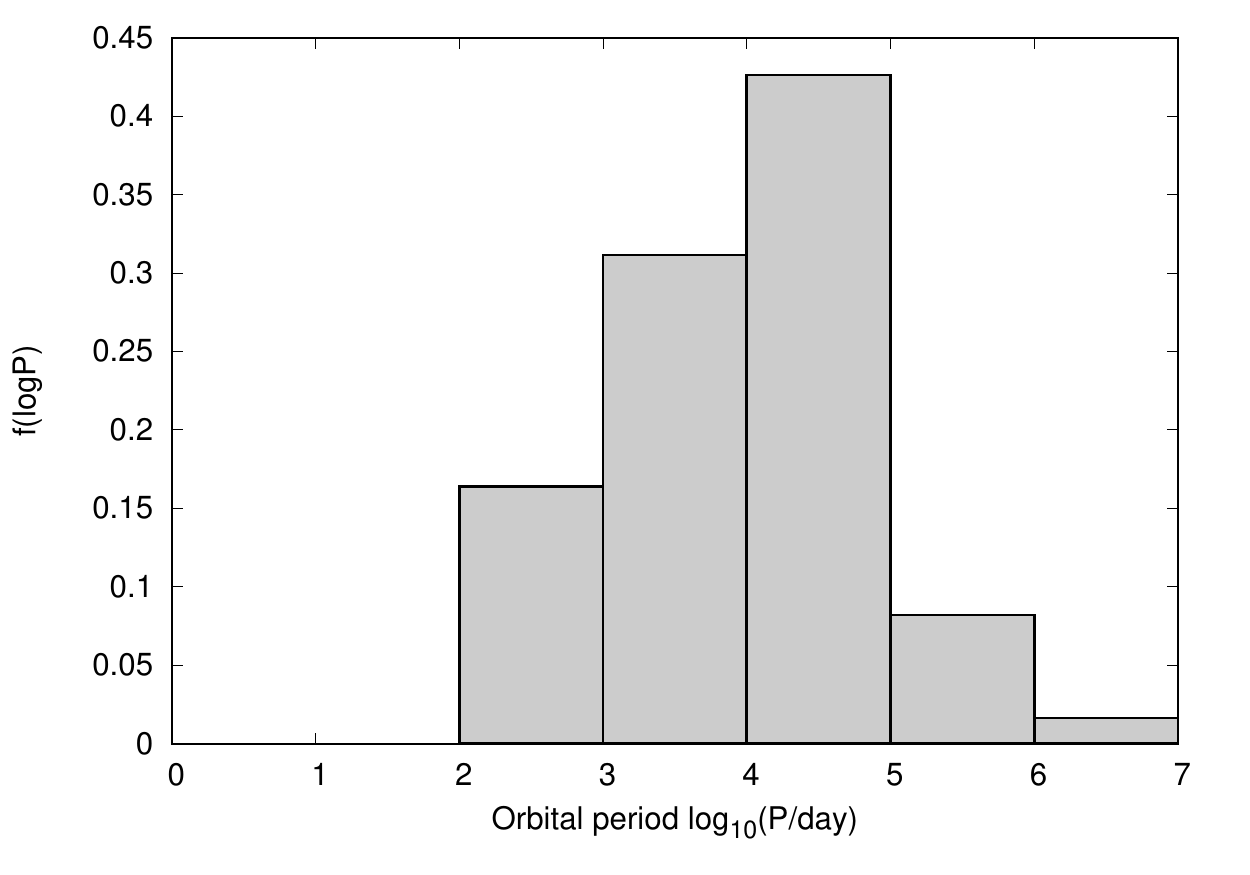}
\includegraphics[width=8.0cm,angle=0]{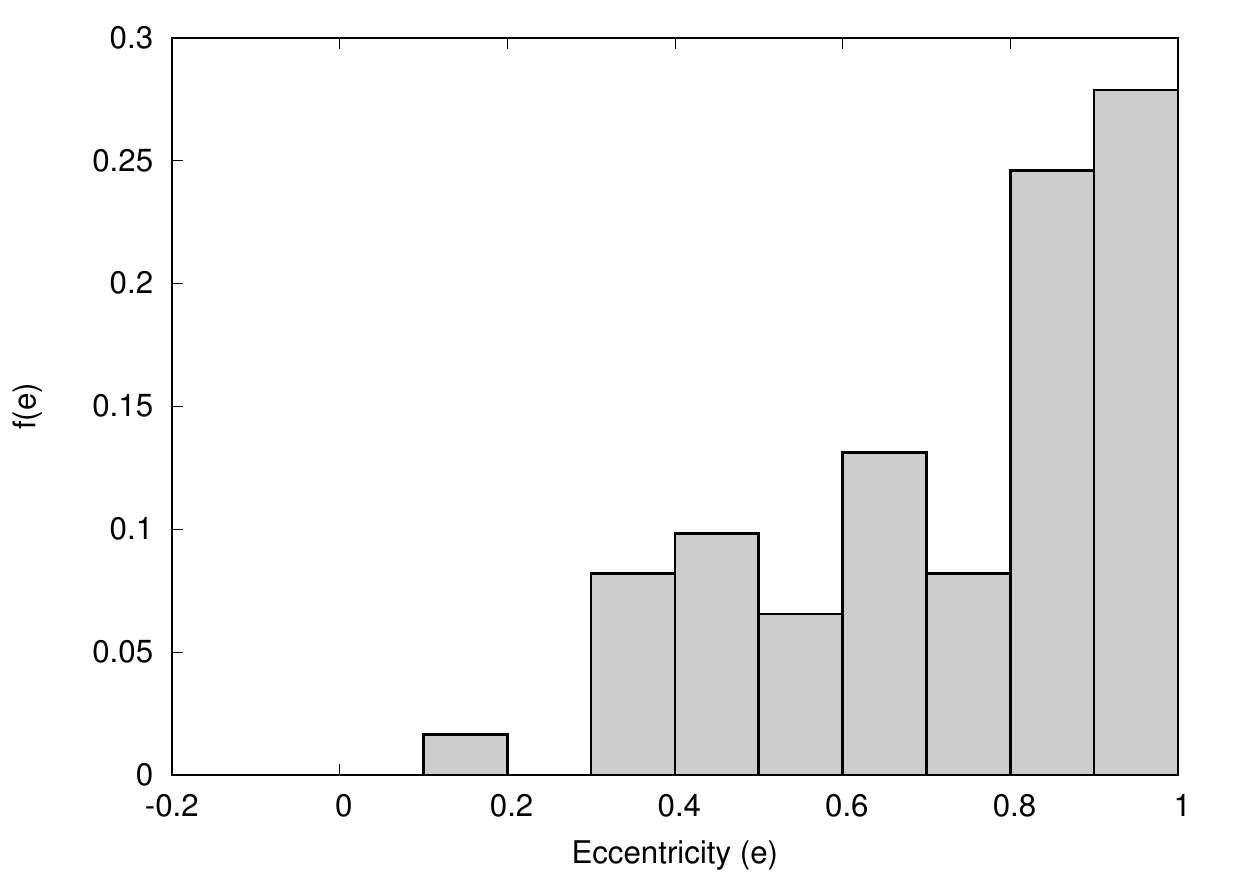}
\includegraphics[width=8.0cm,angle=0]{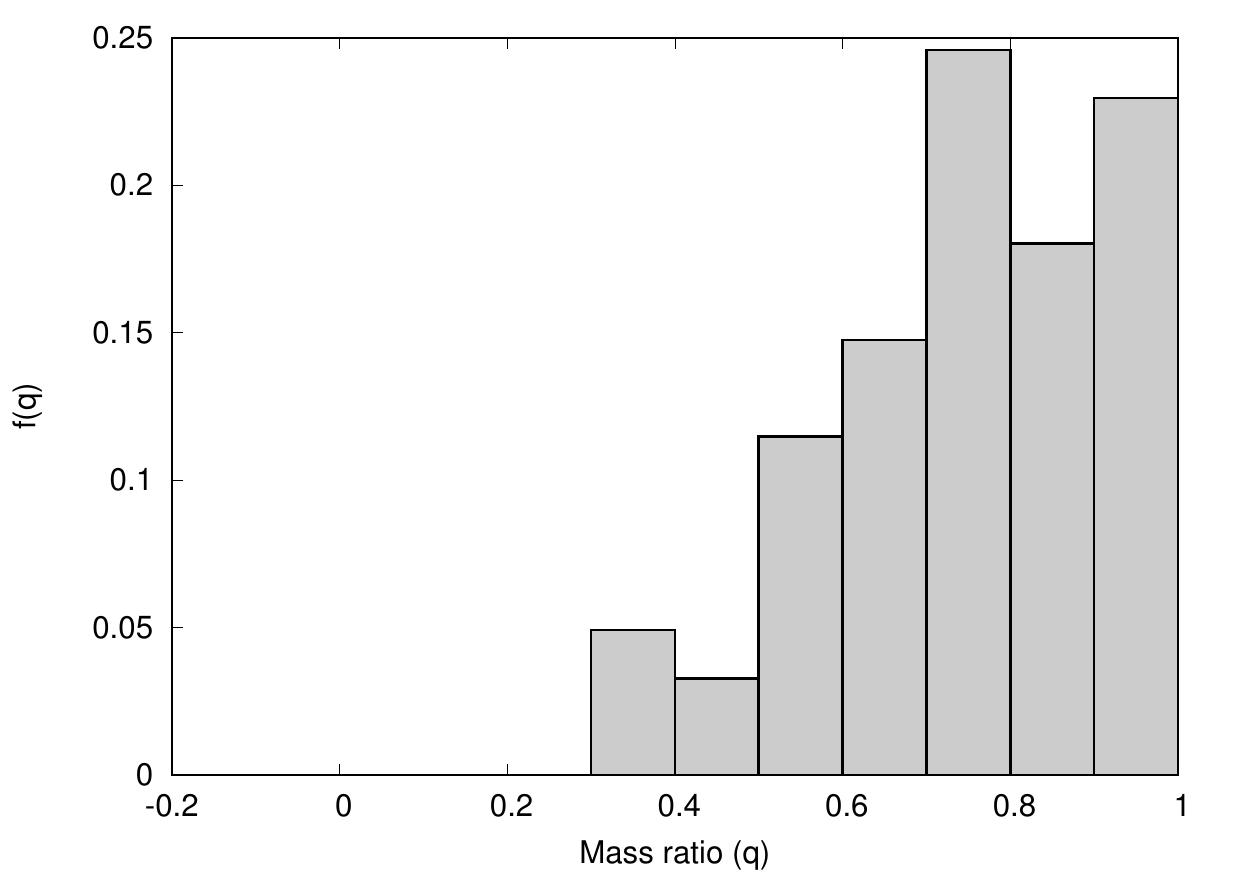}
\caption{The distributions of orbital period (top), eccentricity (center) and
	mass ratio (bottom) of the dynamically-ejected binary black holes (BBHs)
	from all the computed models in Table~\ref{tab:comp}.}
\label{fig:binprop}
\end{figure}

\section{Results}\label{result}

\subsection{The impact of stellar-mass black holes on star cluster evolution}\label{bhdyn}

As outlined in Sec.~\ref{intro}, the primary impact of stellar-mass BH retention in a cluster
just after their formation is to inject energy into the dense stellar environment due to the dynamical
encounters in the BH-core (and also into the BH-core itself). Generally,
for a given mass and compactness of the parent cluster, the energy injection will be more efficient
and consequently the expansion of the cluster will be larger with increasing number of post-birth retained
BHs and as well with increasing mean BH mass. Both of these factors would generally boost the
K.E. generated in dynamical interactions in the BH-core and the energy deposition
onto the stellar environment via dynamical friction
(see Sec.~\ref{intro}; \citealt{2007MNRAS.379L..40M,2008MNRAS.386...65M}).
The strong and frequent dynamical encounters in the
dense BH sub-cluster continue to eject BHs, mainly as single BHs and BBHs.

Fig.~\ref{fig:lagrng_30k}
demonstrates the evolution of $\mcl(0)\approx3\times10^4\Ms$, $\rh(0)\approx2$ pc models with $Z=0.05\Zs$
and $\Zs$. As expected, the low-$Z$ cluster expands by much larger extent than its solar-$Z$ counterpart,
as seen in their Lagrange-radii plots (Fig.~\ref{fig:lagrng_30k}, top row), due to the larger retention
of the BHs at birth (both in number and total mass; Fig.~\ref{fig:lagrng_30k}, middle and bottom rows)
in the former case. The larger rate of expansion and the
correspondingly larger rate of loss of stars across the tidal radius causes the low-$Z$ cluster
to dissolve in $\approx11$ Gyr, while the solar-$Z$ model would survive until the Hubble time. The
low-$Z$ model also continues to retain BHs in larger number and total mass as their dynamical ejections
continue (Fig.~\ref{fig:lagrng_30k}, middle and bottom rows);
see below for more on this point. The larger BH content, in turn, continues to expand the 
$Z=0.05\Zs$ cluster until its point of dissolution, while the $Z=\Zs$ cluster begins to re-collapse
after $\approx4$ Gyr as its BH-engine becomes weak enough due to only a few $\approx10\Ms$ BHs
remaining after this time; in fact the latter cluster undergoes (second) core collapse at $\approx11$ Gyr
(\cf Fig.~\ref{fig:lagrng_30k}, top right panel). On the other hand,
the much lower-mass ECS Ns (of $\approx1.3\Ms$ each; see Sec.~\ref{newwind}), that retain at birth
and are yet to mass-segregate (see below), suffer a larger loss rate in the low-$Z$ case 
(Fig.~\ref{fig:lagrng_30k}, middle and bottom rows) due to the correspondingly faster star removal.

\begin{figure*}
\centering
\includegraphics[width=12.0cm,angle=0]{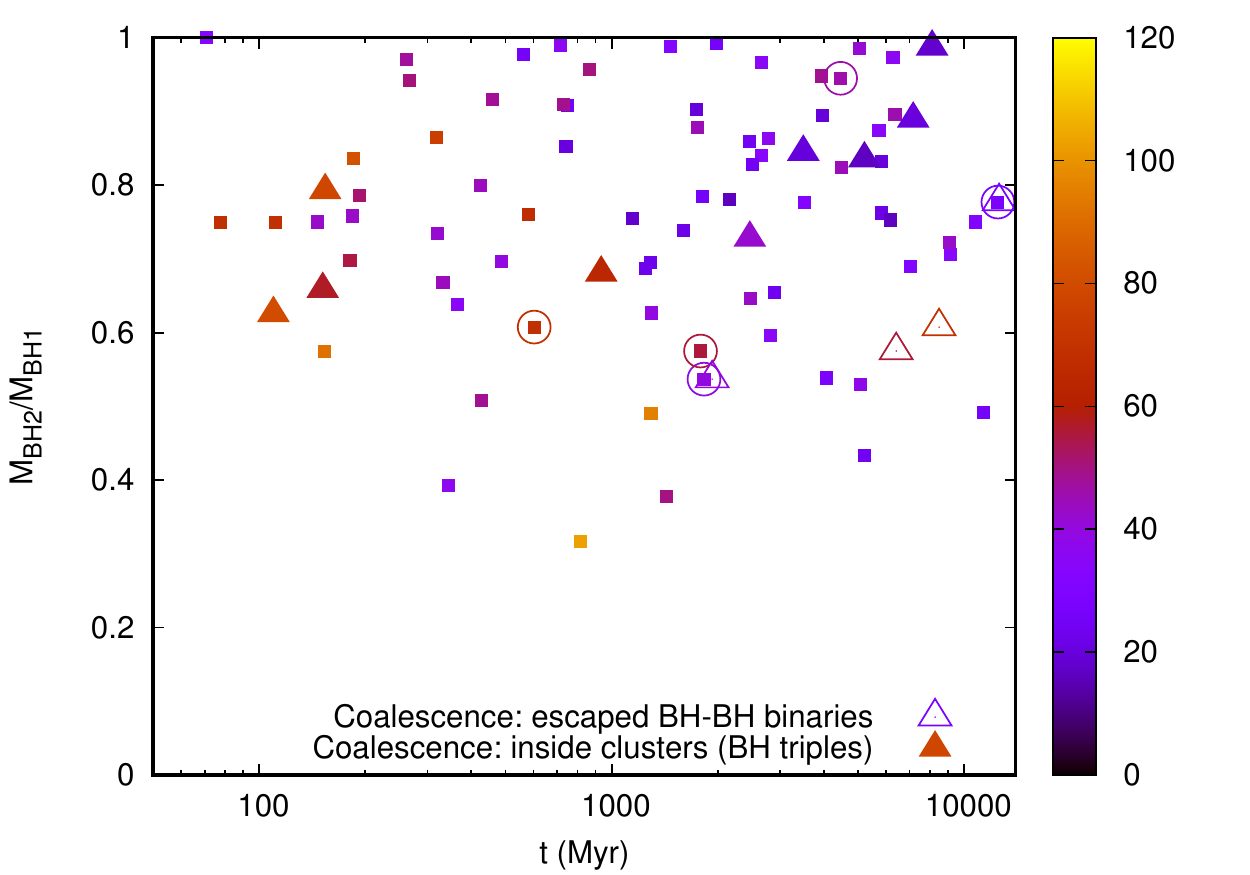}
	\caption{{\bf Filled squares:} the mass ratios of the ejected BBHs against their
	respective cluster-evolutionary time of ejection, $\tej$, from the model clusters.
	{\bf Open circles:} those ejected BBHs with GW coalescence time, $\taumrg\leq13.7$ Gyr (Hubble time)
	at their $\tej$s.
	{\bf Open triangles:} the actual times of GW coalescence, $\tmrg\equiv\tej+\taumrg$,
	of the above BBHs. {\bf Filled triangles:} (triple-mediated) BBH coalescences within the
	clusters at their corresponding coalescence times, $\tmrg$.
	All the symbols are colour-coded according to the BBH's total mass, $\mtot$ (vertical colour bar).
	Results from all the computed models in Table~\ref{tab:comp} are compiled here.}
\label{fig:mrg_ratio}
\end{figure*}

Fig.~\ref{fig:hmr_bhnum_Z} compares the time evolution of the half-mass radius (50\% Lagrange radius), $\rh$,  
and the number, $\nbhbound$, of the BHs bound (\ie, those remaining within the cluster's tidal
radius; except for a few which are on their way of escaping from the cluster, they are also gravitationally
bound to the system) to clusters of fixed $\mcl(0)$ and $\rh(0)$, but of varying $Z$.
As can be expected based on the previous example, with decreasing $Z$, the by-birth retained
$\nbhbound$ and mean BH mass increase, causing the cluster to
expand at a higher rate (Fig.~\ref{fig:hmr_bhnum_Z}, top row).
Also, with decreasing $Z$, an overall larger $\nbhbound$ is maintained;
in fact the $\nbhbound(t)$s for $Z=0.05\Zs$ and $Z=0.25\Zs$ closely follow each other for
most of the time while that for $Z=\Zs$ falls below significantly (Fig.~\ref{fig:hmr_bhnum_Z}, bottom row).
This hints that {\it at least for $Z\lesssim 0.25\Zs$, the number of BHs retaining over time within a cluster of a
given initial mass and size is nearly independent of $Z$. However, for any given observing epoch $t>0$, a lower-$Z$
(tidally under-full) cluster would generally be more expanded.}

It would be also useful to compare $\rh(t)$ and $\nbhbound(t)$, keeping $Z$ and $\rh(0)$ fixed but
varying $\mcl(0)$, as done in Fig.~\ref{fig:hmr_bhnum_M}. Except for $\mcl(0)=5\times10^4\Ms$ and
$Z=0.05\Zs$, which model continues to retain most BHs, the initial expansion of all clusters stall
at some point when their BH-engines become weak enough due to the loss of BHs,
and they begin and continue to contract until their dissolution. Because of generally smaller
number (number and mean mass) of the BH content at all times, the collapse begins earlier with decreasing $\mcl(0)$
(increasing $Z$), for fixed $Z$ ($\mcl(0)$) and $\rh(0)$ (also seen in Fig.~\ref{fig:hmr_bhnum_Z}).
Note that the cluster dissolution times obtained here are exact only for the static and
simplified external field assumed here (Sec.~\ref{runs}), and these times would be different under more realistic
conditions.

\begin{figure}
\centering
\includegraphics[width=8.0cm,angle=0]{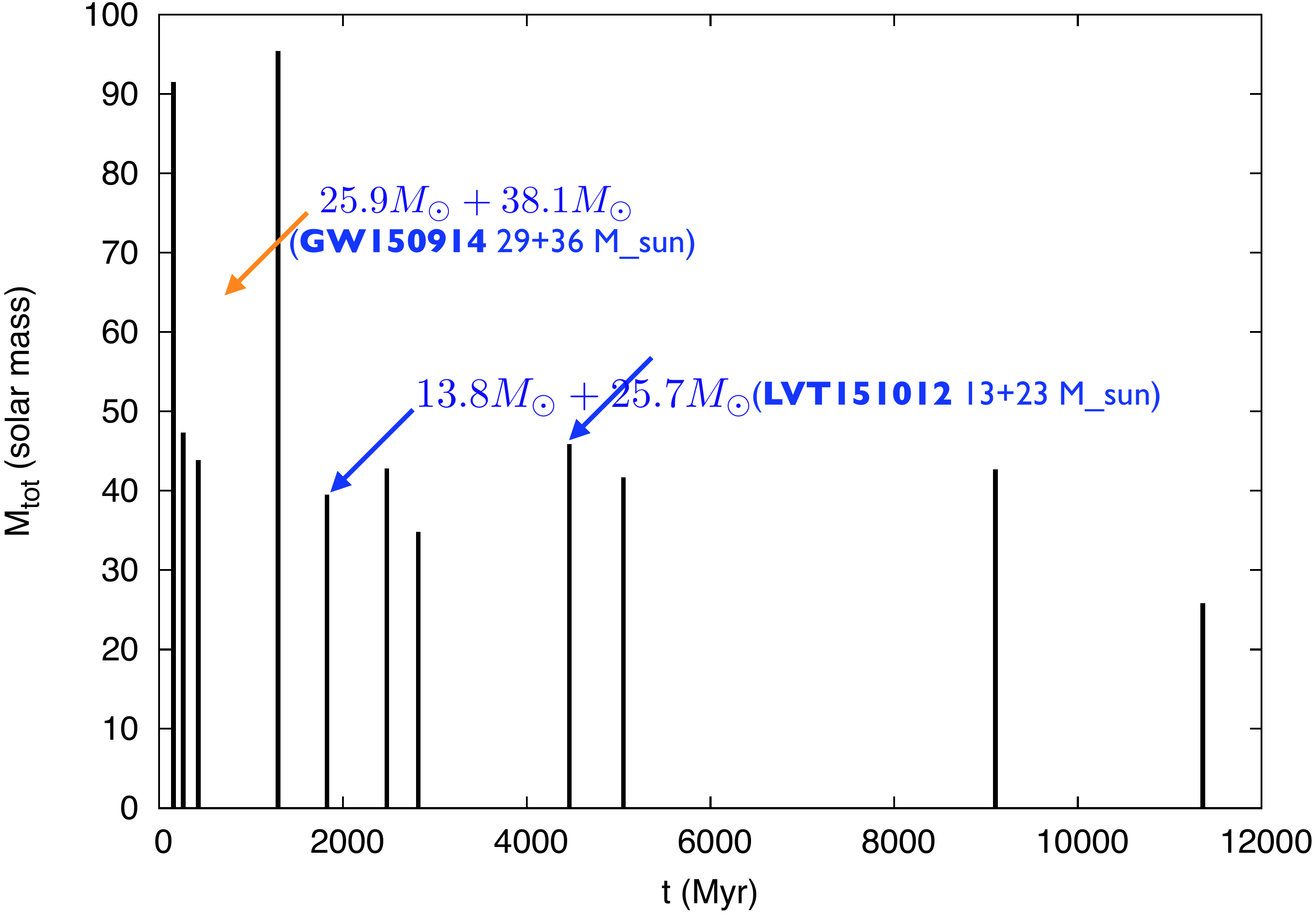}
\includegraphics[width=8.0cm,angle=0]{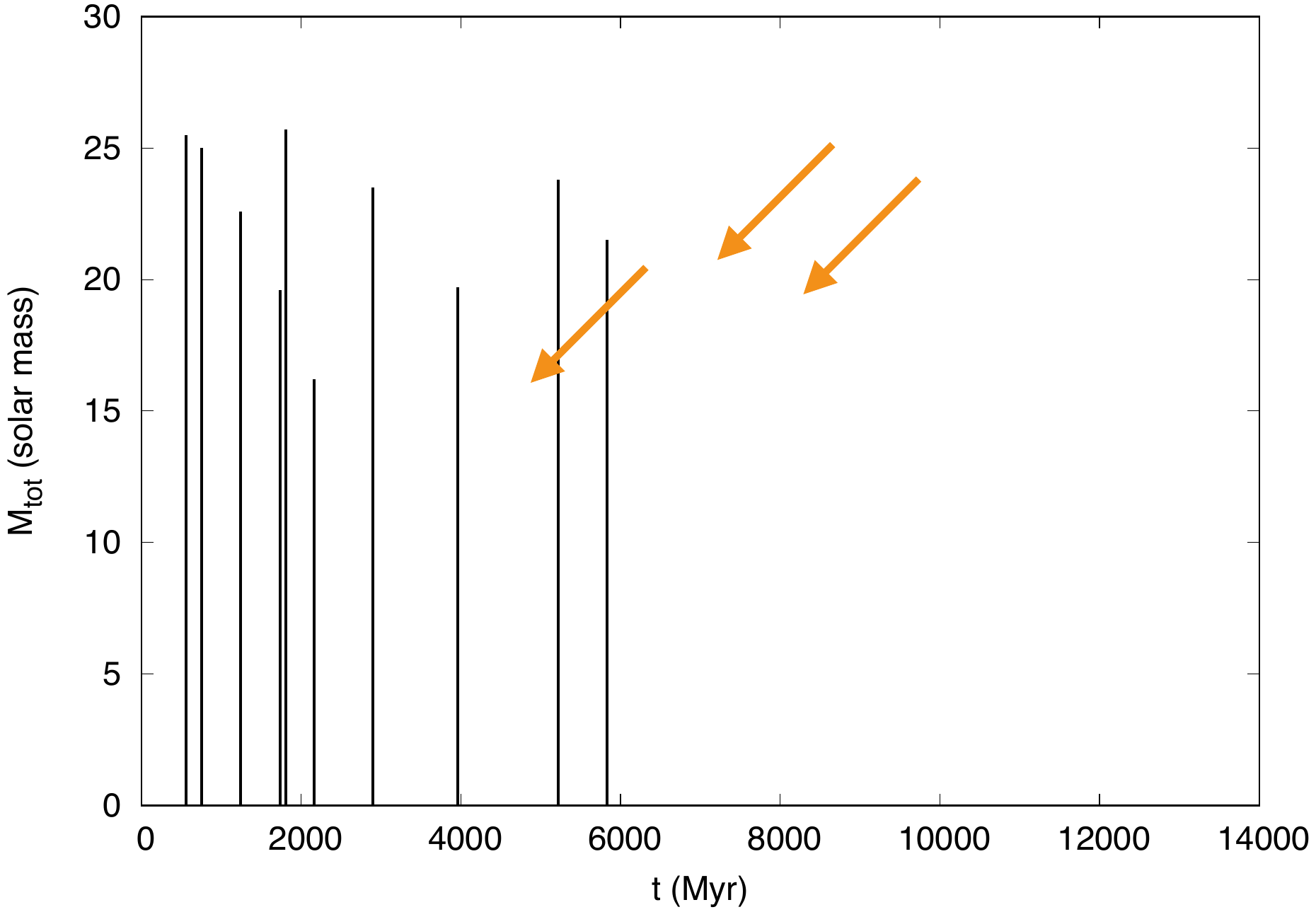}
	\caption{{\bf Top:} the $\mtot$s of the ejected BBHs against their corresponding
	$\tej$s for the computed model with $\mcl(0)\approx3\times10^4\Ms$,
	$\rh(0)\approx2$ pc, $Z=0.05\Zs$. The ejected BBHs with $\taumrg\leq13.7$ Gyr
	are indicated by the blue arrows and the BBH coalescence within
	the cluster is indicated by an orange arrow; \cf Table.~\ref{tab:comp}. The BBH coalescences,
	that resemble the LIGO-detected ones (see Sec.~\ref{intro})
	in terms of their component masses, are indicated. {\bf Bottom:} same
	as the top panel but for the model with $\mcl(0)\approx5\times10^4\Ms$,
	$\rh(0)\approx2$ pc, $Z=\Zs$. All the BBH coalescences occur within
	the cluster, in this particular model; \cf Table.~\ref{tab:comp}.}
\label{fig:BBHesc}
\end{figure}

A relevant question here is that how the {\it efficiency} of the dynamical BH ejection depends on
$\mcl(0)$ and $Z$ (for fixed $\rh(0)$)? This can be best described by plotting the time evolution
of the {\it fraction} of the retaining BHs, as shown in Fig.~\ref{fig:bhfrac}. It can be seen that
for a given $\mcl(0)$ and $\rh(0)$, clusters with $Z\lesssim0.25\Zs$ possess nearly the same efficiency
of BH ejection at all evolutionary times whereas that with $Z=\Zs$ is always more efficient in ejecting
BHs (Fig.~\ref{fig:bhfrac}, left panel)\footnote{In these curves, the decline of $\nbhbound/\nbhboundmx$
after $t\gtrsim100$ Myr is due to the ejections of BHs via dynamical encounters. However, the drop
at $t\sim10$ Myr (distinctly visible due to the use of logarithmic $t$-axis)
is due to the formation of BHs with relatively low natal kicks but which are still unbound
w.r.t. the cluster and are removed shortly when they cross the tidal radius (see above).
These are among the least massive BHs that receive the scaled natal kicks
(Sec.~\ref{newwind}) and which are more numerous for $Z=\Zs$ where the BHs have narrower mass range (Fig.~\ref{fig:bhmass}).
These BHs also cause the excursions of the BH half-mass radius evolution at $t\sim10$ Myr
in Fig.~\ref{fig:hmr_rem_Zsolar} (see below).}.
This is counter-intuitive at the first glance, as lower $Z$ clusters form more massive BHs and
initially retain more of them in number, so that the BH-engine should become more efficient with decreasing
$Z$. This is indeed the case: as demonstrated above, although a lower-$Z$ cluster loses less mass over
its young age from stellar winds and supernovae, it ultimately expands more and dissolves faster
due to the work of its BHs. However, the larger extent of expansion and star loss
causes the normal stellar density to drop faster, reducing the efficiency of dynamical friction
throughout the cluster, and as well diluting the gradient of the central potential well offered
by them. This causes expansion and dilution of the BH-core itself, suppressing the frequency and
K.E. extraction in dynamical encounters within the BH-core, and hence its efficiency. In other words,
\emph{the BH-dynamical heating of a cluster is self-regulatory, suppressing the relative BH ejection rate
for more numerous and/or more massive BH retention at birth.} This is as well manifested
when $Z$ and $\rh(0)$ are kept fixed and $\mcl(0)$ is varied (Fig.~\ref{fig:bhfrac}, right panel).
Note that this self-regulatory behaviour is essentially a manifestation of the \citet{Henon_1975}
principle, according to which the central (dynamical) energy generation of a (post-core-collapse) cluster is controlled
by the energy demands of the bulk of the cluster. This principle is as well
applicable to the energy generation due to dynamical encounters within the BH-core inside a stellar cluster,   
as recently done in the semi-analytic study by \citet{Breen_2013}.
\emph{Nevertheless, the initial retention of more numerous and/or more massive
BH sub-population, ultimately injects more energy (for a given compactness) into
the parent cluster (and onto itself), increasingly prolongigng the BH retention}, as seen from the above examples. For
$\mcl(0)\approx3\times10^4\Ms$ and $5\times10^4\Ms$, the $Z=\Zs$ models are nearly
deprived of their BHs by 10 Gyr whereas the lower-$Z$ models continue to hold
a significant number of BHs until the Hubble time or until the cluster's dissolution;
\cf Fig.~\ref{fig:hmr_bhnum_Z}. This result, therefore, \emph{suggests the presence of a
significant population of BHs in present-day GCs as often envisaged in
the literature (\eg, in \citealt{Strader_2012,Taylor_2015,Bovill_2016,Sollima_2016,Peuten_2016}),
which are typically of sub-solar metallicity,
irrespective of the strength of the external field under which they orbit}.

In the present context, it would be useful to also consider the ECS NSs retaining in the clusters,
since in $\sim$ Gyr old stellar systems they would be the second most massive objects.
Fig.~\ref{fig:hmr_rem_50kZlow} compares the time evolution of the half-mass radius, $\rh(t)$,
of the overall cluster with that of the half-mass radius, $\rhbh(t)$, of the BH subpopulation and
of the half-mass radius, $\rhns(t)$, of the NS subpopulation, for the $\mcl(0)\approx5\times10^4\Ms$,
$\rh(0)\approx2$ pc, $Z=0.05\Zs$ model. After the initial central segregation of the BHs in $\approx100$ Myr,
the BH-core could maintain a fluctuating but overall constant size for a few 100 Myr and then it
expands with the cluster (see discussion above) until the Hubble time, $\rhbh(t)$ being always a
factor of few smaller than $\rh(t)$. This implies a continued centrally-concentrated (or BH-core)
state of the BH sub-population within the cluster, despite significant dilution of both with time.
On the other hand, $\rhns(t)>\rh(t)$ always, especially for $t\gtrsim50$ Myr, which implies lack of mass
segregation of NSs throughout the evolution. The initial large overshoot of $\rhns(t)$ is due to the formation of
NSs via core-collapse supernovae that gives them large kicks (see Sec.~\ref{newwind}) and they
are more likely found much outside the cluster but within the tidal radius ($\sim 100$ pc at that time),
while escaping the system. The low-kick ECS NSs are born during
$t\approx 56.1-65.3$ Myr ($t\approx 40.2-46.0$ Myr) for $Z=0.05\Zs$ ($Z=\Zs$) whose progenitors, of 
$\approx6.8-6.3\Ms$ ($\approx8.2-7.7\Ms$) ZAMS, could not yet fully segregate and, in fact,
the large mass loss until the remnant
formation (the ECS NSs are only of $\approx1.3\Ms$) is likely to reverse any partial segregation achieved
by their progenitors. The central K.E. injection, due to the initial rapid segregation of the retaining BHs and
the continued energetic dynamical encounters among them thereafter (see above), quenches the ``natural''
two-body relaxation driven mass-segregation within the cluster, including that of the NSs. In the
example of Fig.~\ref{fig:hmr_rem_50kZlow}, the mass segregation is frozen until the Hubble time. For
models of higher $Z$, where the BH-engine is weaker and the BHs are depleted relatively fast (see above),
the mass segregation can revive; this is demonstrated in the $Z=\Zs$ examples in
Fig.~\ref{fig:hmr_rem_Zsolar} where the $\rhns(t)$ falls below $\rh(t)$ after a few Gyr of
evolution. Such stalling of mass segregation, due to the work of BHs,
has also been demonstrated in the recent studies by \citet{Alessandrini_2016,Peuten_2016}.

\subsection{Dynamically-formed binary black holes: gravitational-wave coalescence events}\label{bhbin}

\begin{figure*}
\centering
\includegraphics[width=8.0cm,angle=0]{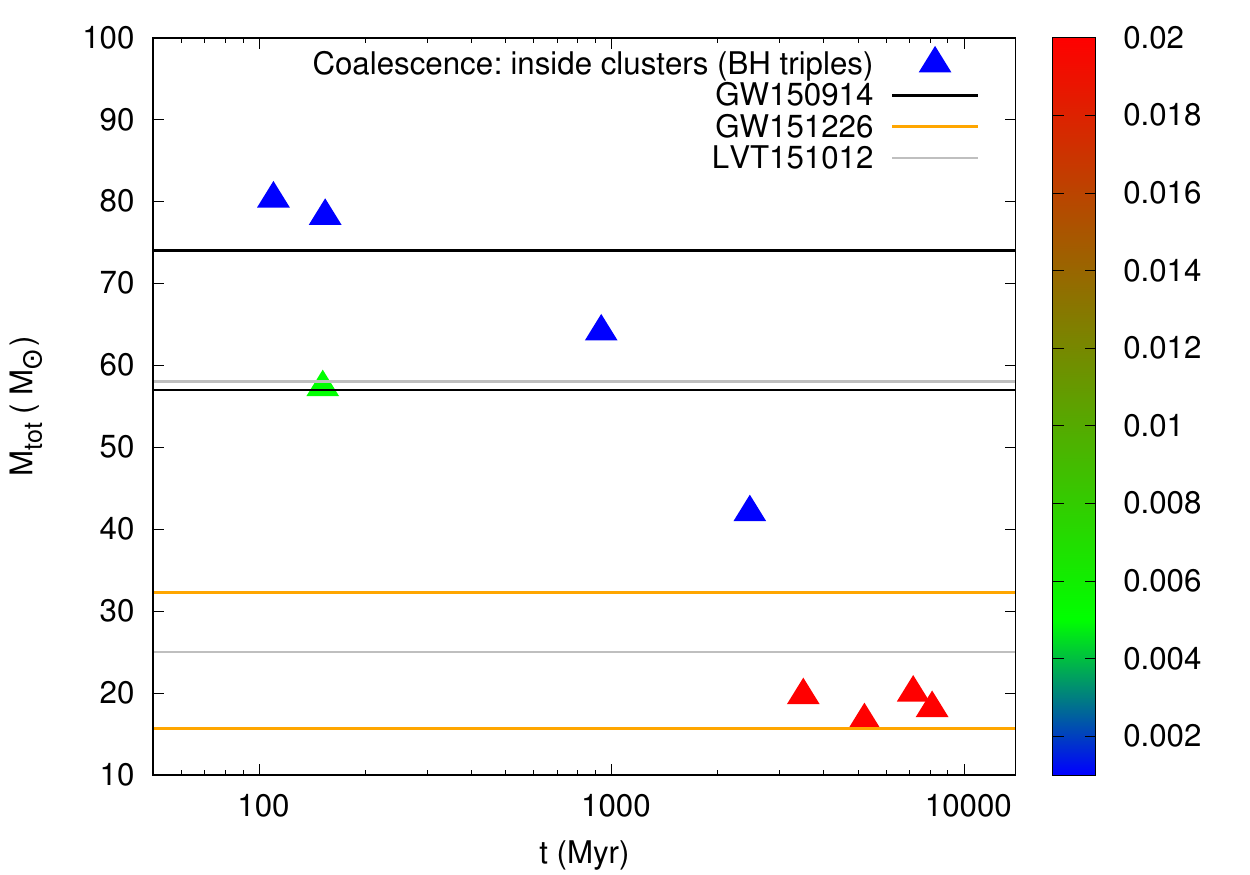}
\includegraphics[width=8.0cm,angle=0]{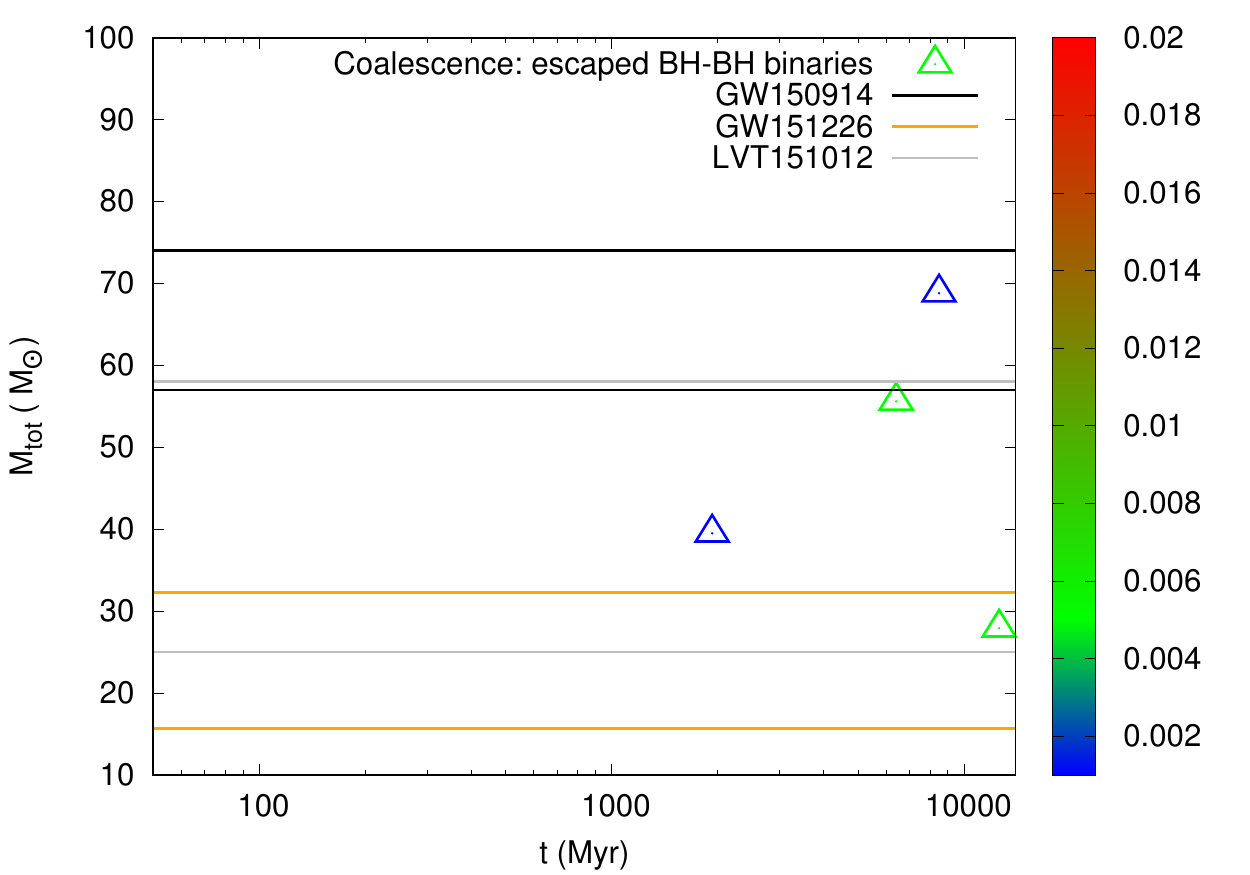}
\includegraphics[width=8.0cm,angle=0]{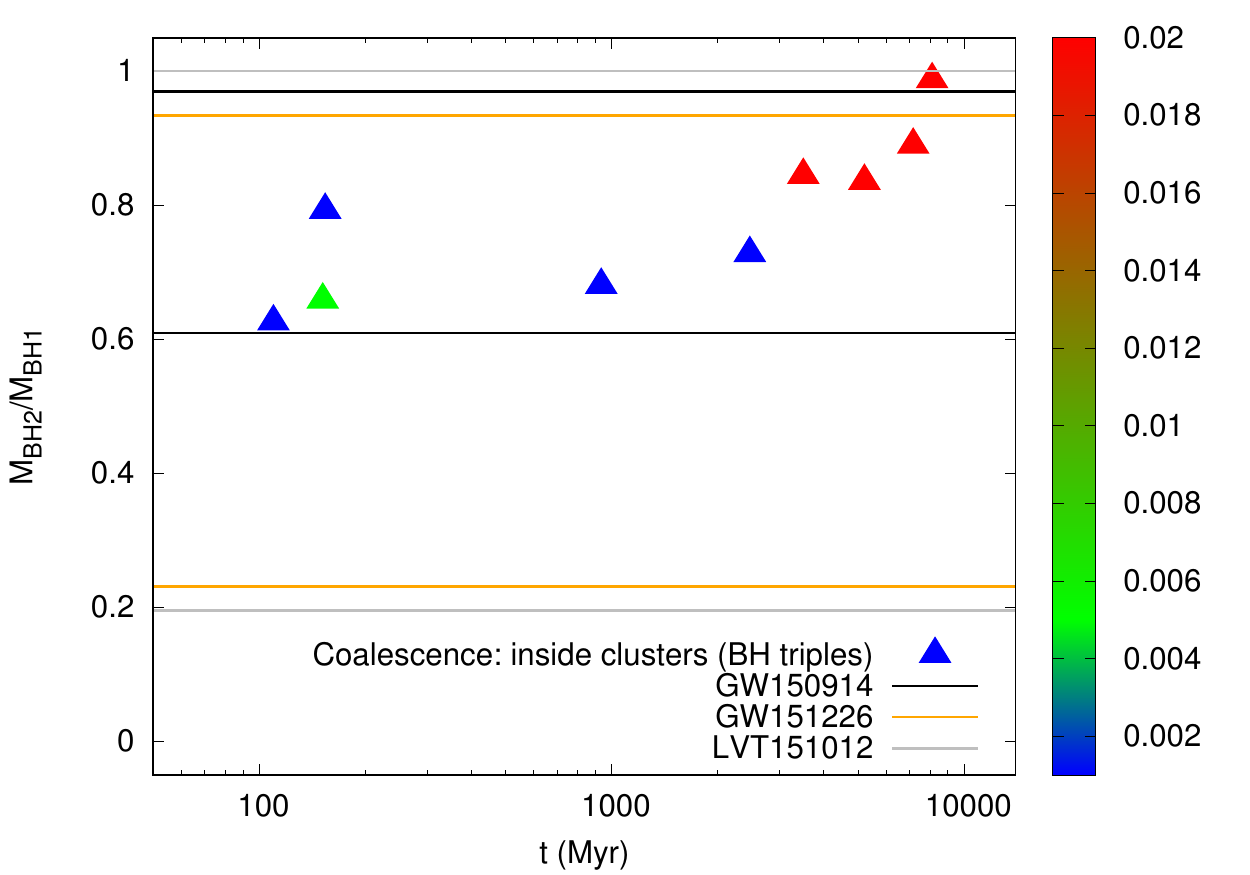}
\includegraphics[width=8.0cm,angle=0]{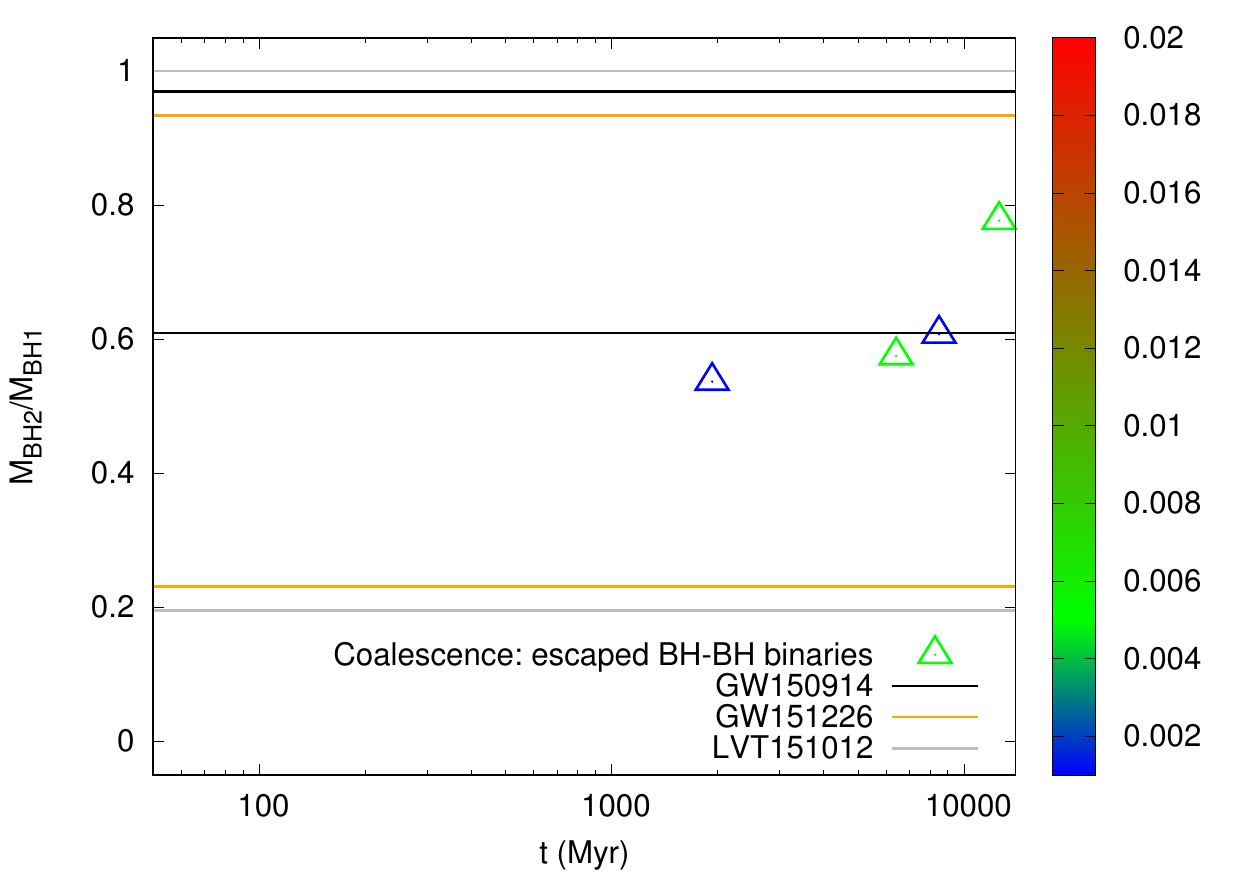}
	\caption{{\bf Top panels:} the $\mtot$s of the in-cluster (triple-mediated; left)
	and the ejected (right) BBH coalescences against their corresponding $\tmrg$s,
	for all the computed models in Table.~\ref{tab:comp}. The colour-coding
	is according to the parent model cluster's $Z$ (vertical colour bar).
        {\bf Bottom panels:} the above BBHs plotted with their mass ratios in the
	Y-axis. In all the panels, the ranges in the Y-axis corresponding to the detected BBH coalescence
	events are indicated by the horizontal lines.}
\label{fig:BBHmrg}
\end{figure*}

Given the current interest in BBH coalescence events following their detection by the LIGO,   
it is undoubtedly worthwhile to investigate such events in the present model computations. This is
of enhanced interest in this work, since YMCs and open-type clusters are dealt with here.
If a power-law shape of the new-born clusters' mass function can be assumed for throughout the Universe,
as observed in our nearby spiral and starburst galaxies
(typically of index $\alpha\approx-2$, \citealt{Gieles_2006a,Gieles_2006b,Larsen_2009}),
such clusters will be the most abundant ones, among those that survive for at least a few Gyr.
Therefore, although expected to be lower than GCs per cluster, their
overall contribution to the observable BBH inspiral rate in the Universe could be at least comparable to
that from classical GCs, thereby potentially adding significantly to the BBH detection
rate from the dynamical channel. This question remains rather unexplored until now.

All the models computed here continue to eject BBHs, which can quench only when 1 or 2 BHs remain bound (in the
$\mcl(0)\approx3\times10^4\Ms$, $Z=\Zs$ model, even the final BH is ejected dynamically
when the central density is increased, as the cluster approached core collapse, although
such complete depletion of BHs is generally unlikely; see Fig.~\ref{fig:lagrng_30k}).
However, the total number of ejected BBHs per cluster is much smaller than that is
typical for more massive Monte-Carlo based models (see Sec.~\ref{intro} and references
therein), as expected. Fig.~\ref{fig:binprop} gives the distributions of orbital period, $P$, eccentricity, $e$,  
and mass ratio, $q$, for the escaped BBHs from all computed models combined. The majority
of these BBHs have $P\sim10^4-10^5$ days\footnote{The majority of the ejected BBHs are from
the $\mcl(0)\approx3\times10^4\Ms$ and $5\times10^4\Ms$ clusters and hence the $P$-distribution
and, particularly, its peak are more of the characteristics of the ejected BBHs from these clusters. Lower-mass
clusters would generally eject wider BBHs and vice versa. The low number of ejected BBHs per cluster here
makes the comparison among the BBH distributions corresponding to different $\mcl(0)$s and $Z$s less meaningful,
which is otherwise advisable.}.
As characteristic of dynamically ejected binaries (via close encounters),
the ejected BBHs are generally of high eccentricity,
with the $e$-distribution peaked beyond $e>0.8$. The mass ratios of the ejected BBHs are
typically of $q>0.5$, with the $q$-distribution function increasing towards $q=1$. This
feature of the $q$-distribution is a signature of the dynamical formation of the BBHs within
the cluster before getting ejected, in which process the pairing of BHs of comparable masses
is energetically favourable.

Except for the least massive ones, all cluster models computed here have produced BBHs
that coalesce within a Hubble time (beginning from the clusters' zero age) due to GW emission,
either still being bound to the cluster (which would typically occur due to Kozai mechanism
in BH-triples; see Sec.~\ref{intro}) or being among the escaped BBHs. The second-last column in
Table.~\ref{tab:comp} shows the number, $\nmrgin$, of BBH coalescences that occurred within each
of the model clusters and also the corresponding component BH masses.
The final column in Table.~\ref{tab:comp} gives, for each model, the number of ejected BBHs,
$\nmrgout$, and their component masses,
that have GW merger time, $\taumrg<13.7$ Gyr (Hubble time), at the time of ejection.
Note that the bound-to-the-cluster coalescences happen in the computations on the fly
(see Sec.~\ref{nbprog}), whereas, for the BBHs ejected from the clusters, the corresponding $\taumrg$s
are estimated using the standard orbit-averaged GW-shrinkage formula by \citet{Peters_1964}.

An interesting fact, that immediately becomes apparent from Table~\ref{tab:comp}, is that, in 
general, $\nmrgin>\nmrgout$, for models with $\mcl(0)>10^4\Ms$. In other words, \emph{YMCs and their
derivative open clusters (as they evolve), are inherently more efficient in producing in-cluster,
triple-mediated BBH coalescences than through ejecting BBHs}.
This is in contrast to what
Monte-Carlo calculations of much more massive systems but with similar model ingredients find (see, \eg,
\citealt{Morscher_2015,Rodriguez_2016,2016arXiv160300884C}). This difference could be due to
artefacts in the Monte-Carlo treatment itself, especially how multiple systems are treated there.
On the other hand, this could as well be characteristic of the lower-mass systems that is
dealt with here; because of lower density of stars and BHs in the present models,
the dynamically-formed triples
can last unperturbed for longer time, giving higher chance to their inner binaries to merge via Kozai
oscillations. This becomes further apparent from the fact that for the models with
$\mcl(0)\leq1.5\times10^4\Ms$, BBH coalescences occur only within the clusters (\cf Table.~\ref{tab:comp}).

Interestingly, such prominence of in-cluster BBH coalescences, as seen here, also contrasts the results
obtained from earlier direct N-body calculations of models of similar mass and size
containing $\approx10\Ms$ BHs, where the escaped BBH coalescences typically dominated
over the in-cluster ones; see, \eg, \citet{2010MNRAS.402..371B}. This is likely to be
the result of a much broader BH mass distribution in the present models. The most
massive couple of BHs would favourably become binary pair within the dense BH-core which
would tend to prevent less massive BHs to pair, and would eject them preferably as singles via
super-elastic scattering. This would suppress the number of ejected BBHs for
a given (initial) mass and size of the cluster, and hence the coalescences among them.
At the same time, the ``bully'' BH-binary, which would typically be the most massive
object in the system, would be harder to eject dynamically, giving it enhanced opportunity
to coalesce (or induce coalescence), if at all, within the cluster. In contrast, in a BH-core
comprising of equal-mass or nearly equal-mass BHs, as in several previous studies,
the dominance of a single BH pair would no more be energetically favourable, which
would reverse the situation. To understand the role of (triple-induced) BBH coalescences
within the cluster, it is necessary to do N-body calculations as in here in larger numbers
and with even higher $\mcl(0)$, which is planned for the near future (see also
\citealt{Kimpson_2016,Haster_2016} in this context).

The filled squares in Fig.~\ref{fig:mrg_ratio}
indicate the mass ratios ($q$s) of the ejected BBHs from all the
models computed here against their respective ejection times ($\tej$s),
which symbols are colour-coded according to their total masses, $\mtot$. The $q$s
corresponding to the in-cluster and after-ejection BBH coalescences are highlighted (by filled and
empty triangles respectively) at the times, $\tmrg$, of their occurrences (therefore,
for the ejected BBH coalescences, $\tmrg\equiv\tej+\taumrg$). As already indicated by Fig.~\ref{fig:binprop}
(bottom panel), most ejected BBHs and all BBH coalescences have $q>0.5$. In-cluster coalescences
can happen from as early as $\tmrg\sim100$ Myr until $\sim10$ Gyr; more massive of those
($60\Ms\lesssim\mtot\lesssim80\Ms$) typically happening within $t\lesssim1$ Gyr. On the other hand,
in the present sample, all coalescences among the ejected BBHs happen after $t\gtrsim1$ Gyr (including
their $\tej$s). Note that the latter conclusion can be an artefact of the low number of
(only 4) ejected coalescences within the Hubble time (see Fig.~\ref{fig:mrg_ratio}); single BHs and BBHs begin
to get ejected as soon as the central BH sub-cluster becomes concentrated enough that three-body
binaries start forming in them (see Sec.~\ref{intro}; this is also when the contraction of
the BH sub-population stalls, see, \eg, Fig.~\ref{fig:hmr_rem_50kZlow}), from $t\sim100$ Myr.

\subsubsection{Comparison with the detected binary black hole coalescences}\label{LIGOcomp}

It would undoubtedly be useful to compare the LIGO-detected events with
the BBH mergers from the present set of computations. Fig.~\ref{fig:BBHesc}
plots the $\mtot$s of the ejected BHHs vs. their $\tej$s for the $\mcl(0)\approx3\times10^4\Ms$,
$\rh(0)\approx2$ pc, $Z=0.05\Zs$ model (top panel) and the
$\mcl(0)\approx5\times10^4\Ms$, $\rh(0)\approx2$ pc, $Z=\Zs$ model (bottom panel).
The ejected BBHs with $\taumrg<13.7$ Gyr are indicated (blue arrows) and as well any coalescences
within the clusters (orange arrows). Remarkably, two mergers in the $\mcl(0)\approx3\times10^4\Ms$
cluster resemble the events GW150914 and LVT151012, both in terms of their $\mtot$s and
also in individual component masses (Sec.~\ref{intro} and references therein).

The $\mcl(0)\approx5\times10^4\Ms$ run is also remarkable in the sense that it has
produced 3 BBH coalescences, all within the cluster and between $5\lesssim t\lesssim8$ Gyr, when nearly 3/4th
of its initially-retained BHs have already escaped (see Fig.~\ref{fig:hmr_bhnum_Z}, lower-right panel).
All the mergers are of GW151226-type, in terms of their $\mtot$s. This
particular calculation is unique among the set in this study, where the lower mass $\approx10\Ms$
BHs (since $Z=\Zs$) allowed the cluster, and hence the BH sub-cluster (see Sec.~\ref{bhdyn}),
to remain sufficiently concentrated throughout its evolution. In particular,
the cluster and its BH-core begin to collapse and boost their concentrations after
$t\approx3$ Gyr (see Fig.~\ref{fig:hmr_rem_Zsolar}, left panel), with a sufficient
number of BHs ($\approx20$) still retaining to continue forming BH-triples which are
relatively uninterrupted. The BHs, in this case, are similarly massive which favour
dynamical BBH formation (see above) and hence the formation of BH-triples within the core.
This implies that \emph{intermediate-aged massive open clusters of solar-like metallicity
serve as highly potential sites for dynamically generating GW151226-like BBH coalescences}.
Similar calculations over a wider parameter range is necessary to reassure this intriguing
inference\footnote{GCs, however, may not be as suitable for creating such a scenario,
due to their much longer two-body
relaxation time, which would cause the re-collapse phase to proceed much slower. However,
they would be more efficient in producing GW151226-like coalescences from ejected BBHs.}.

Fig.~\ref{fig:BBHmrg} shows the $\mtot$s (top row) and $q$s (bottom row) of the in-cluster
(left column) and escaped (right column) BBH coalescences, at their respective $\tmrg$s,
from the present set of computations,
which are compared with the limits of the detected events. All the symbols here are colour-coded
according to the parent cluster's $Z$. As already seen above, the in-situ mergers occur
from age as young as $t\approx100$ Myr up to at least 10 Gyr. Typically, lower $Z$
clusters yield more massive coalescences, which occur at earlier $\tmrg$
(the negative trend in Fig.~\ref{fig:BBHmrg}, top-left panel). This overall trend
is due to the fact that more massive BHs segregate and interact dynamically earlier
(see also \citealt{2016arXiv160906689C} in this context).
The $q$s of all in-situ BBH coalescences are $>0.6$ (Fig.~\ref{fig:BBHmrg}, bottom-left panel),
since pairings within the cluster would preferably happen among
BHs whose masses are close to each other (see above). The lower-mass
BHs, which are dynamically processed later in time, are likely to have partners
that are closer in mass (due to the IMF slope and the BH mass-ZAMS mass relation;
see Sec.~\ref{newwind}), giving rise to the positive trend of the in-situ mergers' $q$s
with $\tmrg$ (Fig.~\ref{fig:BBHmrg}, bottom-left panel). A similar trend follows for
the $q$s of the escaped BBH coalescences; however, all the escaped coalescences happen after
$\approx1$ Gyr (Fig.~\ref{fig:BBHmrg}, bottom-right panel). Also, the escaped coalescences
generally have lower $q$s than their in-cluster counterparts. There is no particular
trend seen in the $\mtot$s of the escaped BBH coalescences (Fig.~\ref{fig:BBHmrg}, top-right panel)
which is likely to be an artefact of the low number of them, in the present sample.

According to Fig.~\ref{fig:BBHmrg}, all of the clusters computed here have the potential
to give rise to BBH coalescence events resembling the detected ones, by dynamical means.
All of the detected events could have occurred either in-situ or after being ejected from their
parent clusters. Finally, at the time of the coalescence, their parent cluster could either
be a $\sim100$ Myr YMC (likely, if the event is in-situ) or be a few - 10 Gyr old
intermediate-mass cluster/open cluster (likely for both in-situ and ejected events).
Interestingly, BBH coalescences, with $\mtot$ exceeding the upper limit of GW150914 by $\approx10\Ms$,  
are also produced in the present computations, as seen in Fig.~\ref{fig:BBHmrg}
(top-left panel). In fact these two in-cluster mergers take place the earliest
($\tmrg\sim100$ Myr) and in the least massive clusters ($\mcl(0)\approx1.0\times10^4\Ms$
and $1.5\times10^4\Ms$; see Table~\ref{tab:comp}). The lower velocity dispersion
in the BH-core of such lower-mass systems would typically produce wider BBHs via three-body
encounters, causing an overall weaker energy extraction in the BH-core and, thereby,
making it harder for the most massive BHs to get ejected (although they begin
to participate in the dynamical interactions the earliest); such clusters
are otherwise most efficient in ejecting BHs, in the sense of Fig.~\ref{fig:bhfrac} (right panel).
This, combined with such clusters' shorter BH-segregation timescale
and shorter two-body relaxation time of the BH-core,
have resulted in such early and massive BBH coalescences. \emph{If BBH coalescences,
substantially more massive than GW150914 (by $\sim10\Ms$), are detected by the Advanced LIGO
in the future, $\sim10^4\Ms$, low-metallicity, compact YMCs would be potential sites for them.}

In making such comparisons, one should bear in mind the cosmological aspects
of it. To detect an event today, one should take into account the time, $\tmrg$,
for the BBH coalescence to occur, counting from the zero age of the parent
cluster, plus time light travel time, $t_L$, for the merger GW signal to reach us from the
location of the event. At its design sensitivity, the Advanced LIGO would detect
the inspiral signal from a pair of $10\Ms$ BHs from maximum $D_{10}\approx1500$ Mpc (comoving radial)
distance (see, \eg, \citealt{2010MNRAS.402..371B}), which corresponds to
a redshift of $z\approx0.37$ and a light travel time of $t_L\approx4.2$
Gyr\footnote{Here we assume $h=0.677$, $\Omega_M=0.309$, flat Universe.} \citep{Wright_2006}.
For a pair of $40\Ms$ BBH coalescence (the most massive ones here; \cf Table~\ref{tab:comp}),
the limiting distance is
$D_{40}\approx4800$ Mpc (Eqns.~6 \& 7 of \citealt{2010MNRAS.402..371B}), corresponding to
$z\approx1.68$ and $t_L\approx9.9$ Gyr. If the trends seen in Fig.~\ref{fig:BBHmrg}
can be bought as a rough representative, then \emph{clusters of all ages can, in principle,
be parents of the detected BBH merger events, depending on the epochs of
their formation}. However, the details of the cosmology
must be taken into account to properly estimate the detection rate of BBH coalescences
(see, \eg, \citealt{Belczynski_2016,2016arXiv160906689C}), which is
beyond the scope of this article.

\subsubsection{LIGO detection rate of binary black hole mergers: a preliminary estimate}\label{LIGOrate}

It would still be useful to make a preliminary estimate of the GW-inspiral detection rate
from dynamically-generated BBH coalescence events, by the LIGO at its design sensitivity,
based on the qualitative aspects learnt from the present computations. We will assume
that all the stellar clusters, contributing to the BBH mergers intercepted at the current epoch,  
are formed $\approx10$ Gyr ago (which is not necessarily true), \ie,
they would currently be either low-$Z$ GCs or low-$Z$ open clusters, if they still survive.
Unlike in \citet{2010MNRAS.402..371B}, where it is assumed that nearly all BHs will
be depleted by $\approx3$ Gyr of cluster evolution and also that most BBH mergers happen
over this time period (based on the results from those calculations), typically a significant
number of BHs continue to retain until the Hubble time or until dissolution, for low-$Z$ systems, as seen in the present
calculations (Sec.~\ref{bhdyn}). Also, except for the least massive systems considered
here, all clusters tend to produce BBH mergers (either in-situ or ejected) over all evolutionary
ages (Sec.~\ref{bhbin}). Hence, not only the intermediate-aged clusters, but also
the classical GCs of the Universe would actively contribute to present-day BBH mergers. Therefore, here we can
take, as the space density of clusters, $\rhocl$,
the sum of those of both GCs and ``young populous clusters'' \citep{2000ApJ...528L..17P}, \ie,
$$\rhocl=11.9 h^3{\rm~Mpc}^{-3},$$
which would contribute to the present-day BBH mergers.

To estimate a bare minimum rate, we will consider the spherical volume of radius $D_{10}$ (see Sec.~\ref{LIGOcomp}),
from within which essentially any BBH coalescence is, in principle, detectable by the LIGO,
at its design sensitivity. If we conservatively assume only one BBH merger event
per cluster, over the corresponding $t_L\approx4.2$ Gyr (see Sec.~\ref{LIGOcomp}), then
the corresponding LIGO detection rate would be $\rligo\approx13\peryr$. Over
$D_{40}$, this rate would scale up to $\rligo\approx425\peryr$. However, the latter rate is an
overestimation since not all BBH inspiral signals can be detected by the LIGO,
from distances beyond $D_{10}$, even at its full sensitivity. Instead, if one modestly assumes
only one BBH inspiral per cluster over 10 Gyr (which time is close to the $t_L$ from $D_{40}$), that can
be detected by the LIGO from within the $D_{40}$ limiting distance, then
$\rligo\approx170\peryr$, at the design sensitivity. Of course, larger contribution
per cluster and contributions from clusters that are formed at later epochs, would
increase both of the limits of $\rligo$. Based on the above preliminary estimates,
it can generally be taken that
$\rligo$, due to dynamical BBH coalescences,
would lie between few 10s to few 100s per year, at the LIGO's design sensitivity.
A more elaborate evaluation of $\rligo$ is planned for the near future.

\section{Summary and outlook}\label{conclude}

The model cluster computations presented in this work, although comprise a preliminary
set but with realistic ingredients (Secs.~\ref{nbprog} \& \ref{newwind}),
provide new and intriguing inferences. They span an initial mass range of
$1.0\times10^4\Ms\lesssim\mcl(0)\lesssim5.0\times10^4\Ms$, a metallicity
range of $0.05\Zs\leq Z \leq\Zs$ and are of half-mass radius $\rh(0)\approx2$ pc;
for models of $\mcl(0)\lesssim1.5\times10^4\Ms$, $\rh(0)\approx1$ pc is also assumed.
All models are evolved by direct N-body method until their dissolution
or otherwise at least for 10 Gyr.
This allows one to study and compare YMC-like systems of varying $Z$,
which evolve into open cluster-like systems, especially w.r.t.
the role of stellar-mass BHs and the dynamically-triggered BBH coalescences.
This mass range has not yet been explored in this way,
although more massive GC-like systems have been studied often (Sec.~\ref{intro}
and references therein).

The BHs that form by direct collapse receive no birth kick and remain
bound to the parent cluster right after their formation (Sec.~\ref{newwind}). The initially-retained
BHs segregate to the cluster center in $\sim 100$ Myr, where they form
a dense BH sub-cluster. The frequent and energetic dynamical encounters
in this BH-engine (or BH-core) continue to inject energy into the dense
normal-stellar bulk of the cluster, until the majority of the BHs are ejected
via the strong dynamical encounters. The energy injection causes the parent
cluster to expand, until the BH-engine is sufficiently weakened due to
the depletion of BHs, after which the cluster begins to re-collapse (Sec.~\ref{bhdyn};
Figs.~\ref{fig:hmr_bhnum_Z} \& \ref{fig:hmr_bhnum_M}). The BH-core also
acts to ``freeze'' the otherwise natural two-body relaxation-driven mass
segregation process (\eg, that of the ECS NSs), which can resume only 
after most of the BHs are dynamically depleted (Figs.~\ref{fig:hmr_rem_50kZlow}
\& \ref{fig:hmr_rem_Zsolar}). Lower-$Z$ clusters
would retain more massive and larger number of BHs at birth (Sec.~\ref{newwind};
Fig.~\ref{fig:bhmass}), which would
have a more profound effect on the cluster. However, the energy injection
process by the BHs seems to be self-regulatory: the more powerful BH-engine
for a lower-$Z$ system also acts to moderate the dynamical BH ejection,
causing a larger fraction of BHs to be retained with time
(Sec.~\ref{bhdyn}; Fig.~\ref{fig:bhfrac}). As a result, the sub-solar-$Z$ clusters
computed here still retain $\sim10$ BHs until the Hubble time or until shortly
before their dissolution, whereas their solar-$Z$ counterparts are nearly
deprived of their BHs by similar evolutionary times (Fig.~\ref{fig:hmr_bhnum_Z}).

All of the computed models of masses $\mcl(0)\approx5\times10^4\Ms$ and
$\approx3\times10^4\Ms$ and the lowest-$Z$ ones of masses
$\mcl(0)\lesssim1.5\times10^4\Ms$ produce BBH coalescences due to inspiral via
GW radiation (Sec.~\ref{bhbin}; Table~\ref{tab:comp}).
They occur either via Kozai mechanism in the dynamically-produced
BH-triples that are bound to the cluster or among the dynamically-ejected eccentric BBHs.
For the present set of models, the majority of the BBH mergers happen triple-mediated,
within the parent cluster. This is in contrast with earlier N-body calculations
that contained BHs of masses similar to each other and also with Monte-Carlo calculations
of more massive clusters having similar model ingredients as the present ones
(see Sec.~\ref{bhbin} and references therein). Given that dynamically-formed
subsystems are treated naturally and accurately in direct N-body calculations
(Sec.~\ref{nbprog}), this is unlikely to be an artefact of the numerical
methods applied here. As explained in Sec.~\ref{bhbin}, a much broader mass spectrum
of the BHs is likely to favour in-situ BBH coalescences, at least over the
mass range of the clusters computed here. It is important to reach even higher
$\mcl(0)$s using the direct N-body method, to properly understand the role of BH-triples
and also the differences with the Monte-Carlo models and with the previous N-body models.
Because of dynamical pairing, most of the BBHs and their coalescences have
mass ratio $q>0.5$ (Sec.~\ref{bhbin}; Fig.~\ref{fig:mrg_ratio}). Among the
coalescences obtained here, there are ones which resemble well with one or other of
the events detected by the LIGO until now (Sec.~\ref{LIGOcomp};
Figs.~\ref{fig:BBHesc} \& \ref{fig:BBHmrg}).
Interestingly, the lowest massive clusters computed here have produced the most
massive and the earliest (in-situ) BBH coalescences, that well exceed the $\mtot$ of GW150914;
this might be a result of weaker dynamical encounters in these systems combined
with their shorter relaxation times (see Sec.~\ref{LIGOcomp}). Again, a larger
number of such lower-mass, low-$Z$ model computations is necessary to ascertain this,
and to understand it better. A back-of-the-envelope conservative estimation is that,
under its proposed design sensitivity, the LIGO should detect $\sim10-\sim100$
BBH inspiral events per year, due to dynamical interactions among stellar-mass BHs
in star clusters alone (Sec.~\ref{LIGOrate}).

The primary uncertainties in the present calculations stem from the
recipes of stellar wind, remnant formation and their natal kicks, that are adopted here
(Sec.~\ref{newwind}). Although such recipes can, in a sense, be called
the state-of-the-art because of their relatively popular applications in
the literature, nearly all
aspects of stellar remnant formation (especially of BH formation; see
Sec.~\ref{newwind}) remain poorly understood or constrained to date.
These uncertainties translate into those in the BHs' mass spectrum and
initial retention in stellar clusters.
Another factor that might have affected the current results to some extent
is the lack of primordial binaries, which is done to make these computations feasible.
Although Monte-Carlo calculations of much more massive clusters indicate
near indifference of the BH dynamics to the presence of primordial binaries (Sec.~\ref{runs}
and references therein),
the situation can be different for lower mass systems, where the
dynamical interactions are generally less energetic. Furthermore, mass transfer and/or
tidal interactions among massive-stellar and close primordial binaries would influence
the masses of the BHs (Thomas Tauris, Philipp Podsiadlowski, private communications;
see also \citealt{DeMink_2009,Marchant_2016} in this context).
Moreover, dynamically induced \citep{Banerjee_2012} or mass-transfer driven \citep{DeMink_2014}
coalescences of massive-stellar
binaries would form more massive merger products, which would yield more massive BHs.

The immediate next step would be to obtain an even more exhaustive set of model
calculations by reaching even higher $\mcl(0)$s and obtaining more evolutionary models
for the lowest cluster masses (and perhaps explore even lower masses).
These will also help to understand better the role of
BH subsystems and their impact on the parent cluster's dynamics and BBH coalescence events,
as discussed above and in the previous sections. Inclusion of primordial binaries
would be feasible at least for the least-massive systems, which would be intriguing to do.
Such steps are planned by the author for the near future.

\section*{Acknowledgements}

The author is indebted to Sverre Aarseth of the Institute of Astronomy, Cambridge,
for his continuous efforts in improving {\tt NBODY6/7}, without which this study wouldn't
have been possible. SB is thankful to Jarrod Hurley of the
Swinburne University of Technology for reviewing the changes made in the {\tt BSE}'s
routines for this work, testing them independently, and also for interesting discussions.
SB is thankful to Chris Belczynski of the Astronomical Observatory, University of Warsaw,
for supplying the data corresponding to the ZAMS mass-remnant mass relations,
as obtained from the {\tt StarTrack} program (Fig.~\ref{fig:bhmass}),
and also for useful discussions. SB is thankful to the anonymous referee
for helpful comments that improved some of the descriptions of the paper. Finally,
but not the least, SB is indebted to the computing team of the Argelander-Institut f\"ur Astronomie,
University of Bonn, for their support and efficient maintenance of the workstations
on which all the computations have been performed.

%\nocite{*}

\bibliographystyle{mnras}
\bibliography{bibliography/biblio.bib}

\label{lastpage}
\end{document}